\documentclass[twocolumn,floatfix]{revtex4}
\pdfoutput=1
\usepackage{graphicx}
\usepackage{dcolumn}
\usepackage{longtable}
\usepackage{amsmath}
\usepackage{amssymb}

\usepackage{braket}
\usepackage{latexsym}
\usepackage{adjustbox}
\usepackage{lscape}

\usepackage{pdflscape}

\usepackage{tikz}
\usetikzlibrary{positioning,shapes,fit,arrows}

\definecolor{myblue}{RGB}{56,94,141}

\newcommand{\abs}[1]{\left\lvert#1\right\rvert}
\begin{document}
\title{Proxy-SU(3) symmetry in the shell model basis
}

\author
{Andriana Martinou$^1$, Dennis Bonatsos$^1$, N. Minkov$^2$, I.E. Assimakis$^1$, S. K. Peroulis$^1$  S. Sarantopoulou$^1$, and J. Cseh$^3$}

\affiliation
{$^1$Institute of Nuclear and Particle Physics, National Centre for Scientific Research 
``Demokritos'', GR-15310 Aghia Paraskevi, Attiki, Greece}

\affiliation
{$^2$Institute of Nuclear Research and Nuclear Energy, Bulgarian Academy of Sciences, 72 Tzarigrad Road, 1784 Sofia, Bulgaria}

\affiliation
{$^3$Institute for Nuclear Research, Debrecen, Pf. 51, Hungary-4001}

\begin{abstract}

The proxy-SU(3) symmetry has been proposed for spin-orbit like nuclear shells using the asymptotic deformed oscillator basis for the single particle orbitals, in which the restoration of the symmetry of the harmonic oscillator shells is achieved by a change of the number of quanta in the $z$-direction by one unit for the intruder parity orbitals. The same definition suffices within the cartesian basis of the Elliott SU(3) model. Through a mapping of the cartesian Elliott basis onto the spherical shell model basis, we translate the proxy-SU(3) approximation into spherical coordinates, proving, that in the spherical shell model basis the proxy-SU(3) approximation corresponds  to the replacement of the intruder parity orbitals by their de Shalit--Goldhaber partners. Furthermore it is shown, that the proxy-SU(3) approximation in the cartesian Elliott basis is equivalent to a unitary transformation in the 
$z$-coordinate, leaving the $x$-$y$ plane intact, a result which in the asymptotic deformed oscillator coordinates implies, that the $z$-projections of angular momenta and spin remain unchanged. The present work offers a microscopic justification of the proxy-SU(3) approximation and in addition paves the way, for taking advantage of the proxy-SU(3) symmetry in shell model calculations. 

 \end{abstract}


\maketitle

\section{Introduction}

In the era of supercomputers symmetries continue, to play an indispensable role in nuclear physics. First they can provide predictions for the overall behavior of measurable quantities and the selection rules they are forced, to obey \cite{IA,IVI,FVI,Rosensteel,RW,Kota}. Second they offer extremely effective shortcuts, largely reducing the size of microscopic calculations \cite{Launey1,Launey2}. While the spherical shell model \cite{Mayer1,MJ} remains the basic microscopic theory of atomic nuclei explaining the appearance of the experimentally observed magic numbers, the existence of the SU(3) symmetry, pointed out by Elliott \cite{Elliott1,Elliott2,Elliott3} in the $sd$ shell, has shown, how quadrupole deformation occurs within a nuclear shell bearing the overall U(6) symmetry of the relevant harmonic oscillator shell \cite{Wybourne,Smirnov,IacLie,BK}, possessing an SU(3) subalgebra. Furthermore the SU(3) subalgebra offers a classification scheme for all nuclei in the said shell.  

The recently introduced proxy-SU(3) symmetry \cite{proxy1,proxy2} is an approximation scheme, which extends the validity of the Elliott SU(3) theory to higher nuclear shells allowing one, to take advantage of the computational tools provided by SU(3) \cite{Kota} for calculations in heavy nuclei away from closed shells, where microscopic calculations still are of prohibitive size. The proxy-SU(3) symmetry has been found \cite{proxy2}, to provide parameter-independent predictions for the collective deformation variables $\beta$ and $\gamma$ \cite{BM}, measuring the deviation from the spherical shape and from axial symmetry respectively. As by-products the dominance \cite{Hamamoto} of prolate over oblate shapes in the ground states of even-even nuclei, as well as the location of a prolate to oblate transition \cite{Linnemann} in heavy deformed nuclei have been obtained \cite{proxy2,proxy3}. 

Within nuclear shells above the $sd$ shell the SU(3) symmetry is known, to be broken by the spin-orbit interaction \cite{Mayer1,MJ}, which in each shell pushes down to the shell below the orbitals bearing the highest total angular momentum $j$. For example in the $sdg$ shell, containing the orbitals $3s^{1/2}$, $2d^{3/2}$, $2d^{5/2}$, $1g^{7/2}$, $1g^{9/2}$, the $1g^{9/2}$ orbital is pushed into the $pf$ shell below, while the $1h^{11/2}$ is coming into the $sdg$ shell from the $pfh$ shell above, called the abnormal parity orbital, since it bears parity opposite to that of the orbitals of the $sdg$ shell, called in this case the normal parity orbitals. 
Within the proxy-SU(3) scheme \cite{proxy1,proxy2} the SU(3) symmetry is restored by replacing the levels of the $1h^{11/2}$ orbital (except the two with the projections 
$m_j=\pm j$ of the total angular momentum $j$) by the levels of the $1g^{9/2}$ orbital with the same projection of the total angular momentum $m_j$.  

The validity of the proxy-SU(3) symmetry has been discussed so far in the framework of the Nilsson model \cite{Nilsson1,NR}, which provides a simple description of deformed nuclei in terms of an axially deformed harmonic oscillator including a spin-orbit interaction and an $l^2$ term (where $l$ is the orbital angular momentum) flattening the bottom of the potential.  The replacement is motivated by the experimentally verified observation \cite{Burcu2010}, that pairs of orbitals differing by $\Delta K [\Delta \mathcal{N} \Delta n_z \Delta \Lambda]=0[110]$ in the Nilsson notation, where $\mathcal{N}$ is the total number of oscillator quanta,  $n_z$ is the number of oscillator quanta along the $z$-axis, $\Lambda=\abs{m_l}$ is the absolute value of the projection of the orbital angular momentum and $K$ is the projection of the total angular momentum along the $z$-axis, are known to exhibit maximal spatial overlaps \cite{Sofia2013}. It has been proved \cite{proxy1}, that the modifications inflicted by these replacements in the matrix elements of the Nilsson Hamiltonian are minimal, thus the single-particle spectra constituting the Nilsson diagrams remain nearly unaltered. 

In the present work we consider the proxy-SU(3) approximation in the framework of the spherical shell model, motivated by the following points and questions. 

a) In the framework of the spherical shell model the usual three-dimensional harmonic oscillator (3D-HO) shells are involved having $U(\Omega)$ symmetries (where $\Omega={(\mathcal{N}+1)(\mathcal{N}+2)/ 2}$), possessing SU(3) subalgebras \cite{Wybourne,Smirnov,IacLie,BK}, while in the Nilsson model deformed oscillator algebras are involved bringing in extra mathematical complications \cite{Asherova,Lenis}, which in the present approach we are going to avoid. 

b) In order to use the proxy-SU(3) approximation within the spherical shell model, one should know, which rule to use for the replacement of orbitals,  to restore the SU(3) symmetry. In the Nilsson notation the appropriate replacements involve the above mentioned $0[110]$ pairs. What kind of pairs should be used in the notation of the spherical shell model?

c) Alternative approximate SU(3) schemes are provided by the pseudo-SU(3) symmetry \cite{Adler,Shimizu,pseudo1,pseudo2,DW1,DW2,Harwood,Ginocchio1,Ginocchio2}, to be further discussed in Section 8, as well as by the quasi-SU(3) scheme \cite{Zuker1,Zuker2}. In the case of the pseudo-SU(3) scheme in a given nuclear shell the normal parity orbitals are replaced by pseudo-SU(3) counterparts, to which they are connected through a unitary transformation \cite{AnnArbor,Quesne,Hess}. Is there any unitary transformation connecting, the orbitals replaced within the proxy-SU(3) framework?   

In short, in this manuscript the spherical shell model justification of the proxy-SU(3) symmetry is offered, based on two mathematical pillars, a transformation from the cartesian Elliott basis to the spherical shell model basis, along with a unitary transformation within the Elliott basis, and a physical pillar, the de Shalit--Goldhaber pairs \cite{deShalit}, to be discussed later. 

In Section 2 of the present work we are going to study the transformation connecting the cartesian Elliott basis to the spherical shell model basis, while in Section 3 the de Shalit--Goldhaber nucleon pairs \cite{deShalit} are going to be studied within this mathematical framework. The importance of the $x$-$y$ plane in the case of axially deformed nuclei will be discussed in Section 4, while in Section 5 we are going to show, that the replacements needed, in order to employ the proxy-SU(3) scheme within the spherical shell model, correspond to de Shalit--Goldhaber nucleon pairs \cite{deShalit}. A unitary transformation connecting, the orbitals involved in the proxy-SU(3) approximation, will be shown, to exist in Sections 6 and 7 and will be compared to the unitary transformation, used within the pseudo-SU(3) scheme in Section 8, while Section 9 
will contain discussion of the preset findings and plans for further work. 

\section{Transformation among the spherical shell model and the cartesian Elliott basis}\label{Trans}

 The simplest single particle Hamiltonian for one nucleon in the atomic nucleus is \cite{Mayer1,MJ}
\begin{equation}\label{H}
H={\mathbf{p}^2\over 2M}+{1\over 2}M\omega^2 \mathbf{r}^2+V_{ls} \mathbf{l}\cdot \mathbf{s},
\end{equation}
where the first two terms represent the three dimensional isotropic harmonic oscillator, with $\mathbf{p}$, $\mathbf{r}$, $M$, $\omega$ being the momentum, spatial coordinate, mass and oscillation frequency respectively, while the last term is the spin-orbit interaction, with $\mathbf{l}$, $\mathbf{s}$ being the orbital angular momentum and spin, and $V_{ls}$ is a strength parameter of the spin-orbit coupling. The spatial, single particle states of the Elliott SU(3) symmetry \cite{Elliott1,Elliott2,Elliott3} are the solutions of the 3D isotropic HO in the cartesian coordinate system $\ket{n_z,n_x,n_y}$, which are Hermite polynomials \cite{Cohen}. The 3D isotropic HO can also be solved in the spherical coordinate system, which is preferred in the shell model \cite{Mayer1,MJ}, giving single particle states $\ket{n,l,m_l}$ \cite{Cohen}, where the total number of oscillator quanta $\mathcal{N}$ is
\begin{equation}\label{N}
\mathcal{N}=2n+l=n_z+n_x+n_y,
\end{equation}
with $m_l$ being the projection of the orbital angular momentum on the $z$-axis, while $n$ is the radial quantum number getting values $n=0,1,2,3,...$

A unitary transformation between the cartesian\\ $\ket{n_z,n_x,n_y}$ and spherical basis $\ket{n,l,m_l}$ is possible
\begin{equation}
\ket{n_z,n_x,n_y}=    \sum_{n,l,m_l}\braket{n,l,m_l|n_z,n_x,n_y}\ket{n,l,m_l},
\end{equation}
with $\mathcal{N}$ and $n$ given by Eq. (\ref{N}),  while $l=\mathcal{N}$, $\mathcal{N}-2$, \dots, 1 or 0,  and $-l\le m_l\le l$. For a given oscillator shell with number of quanta $\mathcal{N}$, the number of degenerate states is ${(\mathcal{N}+1)(\mathcal{N}+2)\over2}$. We have doubled these states, {\it i.e.}, we have encountered each $\ket{n_z,n_x,n_y}$ and $\ket{n,l,m_l}$ orbital twice in the basis, since two identical nucleons with opposite spin projections ($\pm1/2$) will be allowed by the Pauli principle, to occupy the same spatial orbital after the inclusion of the spin. This step was necessary, in order to be able, to multiply the matrix $R$ with the matrix $C$ in Eq. (\ref{RS}). The doubled cartesian states $\ket{n_z,n_x,n_y}$ are the matrix elements of a column matrix $(\mathcal{N}+1)(\mathcal{N}+2)\times 1$ labeled as $[n_zn_xn_y]$, while  the relevant column matrix, which consists of the doubled spherical states $\ket{n,l,m_l}$, is labeled as $[nlm_l]$. While $R$ was originally of dimension ${(\mathcal{N}+1)(\mathcal{N}+2)\over 2}\times {(\mathcal{N}+1)(\mathcal{N}+2)\over 2}$, after the doubling of the states $R$ is the square $(\mathcal{N}+1)(\mathcal{N}+2)\times (\mathcal{N}+1)(\mathcal{N}+2)$ transformation matrix among the two bases and it holds, that:
\begin{equation}\label{R}
[n_zn_xn_y]=R\cdot [nlm_l].
\end{equation}
 The matrix elements of the transformation matrix $R$ for $m_l\ge 0$ are given by \cite{Chacon,Davies,Chasman1967}
\begin{eqnarray}
\braket{n,l,m_l|n_z,n_x,n_y}=\delta_{2n+l,n_x+n_y+n_z}\nonumber\\(-1)^{(2n+n_x+n_y-m_l)/2}\cdot i^{n_y}\nonumber\\\left({(2l+1)(l-m_l)!(n+l)!\over 2^l(l+m_l)!n!(2n+2l+1)!} \right)^{1/2}
\left({n_x+n_y+m_l\over2}\right)!\nonumber\\(n_x!n_y!n_z!)^{1/2} \left({1+(-1)^{n_x+n_y+m_l}\over 2} \right)\nonumber\\\sum_{t=t_{min}}^{t_{max}}{(-1)^t(2l-2t)!(n+t)!\over t!(l-t)!(l-2t-m_l)!(n+t-{n_x+n_y-m_l\over 2})!}\nonumber\\
\sum_{h=h_{min}}^{h_{max}}{(-1)^h\over h!(n_x-h)!(h+{n_y-n_x-m_l\over 2})!({n_x+n_y+m_l\over 2}-h)!},
\end{eqnarray}
where
\begin{eqnarray}
t_{min}=\begin{cases}
0,& \mbox{ for } n\ge {n_x+n_y-m_l\over 2}\\
{n_x+n_y-m_l\over 2}-n, &\mbox{ for } n< {n_x+n_y-m_l\over 2}
\end{cases},\\
t_{max}=\begin{cases}
{l-m_l\over 2}, & \mbox{ if } l-m_l\mbox{ is even}\\
{l-m_l-1\over 2}, & \mbox{ if }l-m_l\mbox{ is odd}
\end{cases},\\
h_{min}=\begin{cases}
0,&\mbox{ for }n_y\ge n_x+m_l,\\
{n_x+m_l-n_y\over 2},&\mbox{ for }n_y< n_x+m_l
\end{cases},\\
h_{max}=\begin{cases}
n_x,&\mbox{ for }n_x\le n_y+m_l,\\
{n_x+n_y+m_l\over 2},&\mbox{ for }n_x> n_y+m_l
\end{cases}.
\end{eqnarray}
For $m_l<0$ it is valid that \cite{Chacon}
\begin{equation}
\braket{n,l,-m_l|n_z,n_x,n_y}=(-1)^{n_x}\braket{n,l,m_l|n_z,n_x,n_y}.
\end{equation}

If the spinor of each nucleon is $\ket{s,m_s}$, with $s=1/2$ and $m_s=\pm 1/2$ being the spin projection, then the spin-orbit coupling $\mathbf{l}\cdot \mathbf{s}$ yields, that the total angular momentum $\mathbf{j}$ is $\mathbf{j}=\mathbf{l}+\mathbf{s}$. The shell model orbitals $\ket{n,l,j,m_j}$ arise after the 
$\mathbf{l}\cdot \mathbf{s}$ coupling, with $m_j=m_l+m_s$ being the projection of the total angular momentum. The $\ket{n,l,j,m_j}$ correspond to the usual shell model notation if one adds 1 unit in the radial quantum number $n$ and represents the angular momentum $l=0,1,2,...$ by the small latin characters $s$, $p$, $d$, \dots, according to the spectroscopic notation. For instance the orbitals $\ket{n,l,j,m_j}$: $\ket{0,1,{3\over2},{1\over 2}}$, $\ket{1,2,{5\over 2},{3\over 2}}$ are labeled $1p^{j=3/2}_{m_j=1/2}$, $2d^{j=5/2}_{m_j=3/2}$ in the shell model notation respectively. The transformation among the shell model basis $\ket{n,l,j,m_j}$ and the spherical basis $\ket{n,l,m_l}\ket{s,m_s}=\ket{n,l,m_l,m_s}$ is achieved through the Clebsch-Gordan (CG) coefficients
\begin{equation}\label{CG}
\ket{n,l,m_l,m_s}=\sum_{m_l,m_s} C^{l s j}_{m_l m_s m_j}\ket{n,l,j,m_j}.
\end{equation}
The CG coefficients form a square $(\mathcal{N}+1)(\mathcal{N}+2)\times (\mathcal{N}+1)(\mathcal{N}+2)$ matrix $C$. Therefore Eq. (\ref{CG}) is written in matrix representation
\begin{equation}\label{C}
[nlm_lm_s]=C\cdot [nljm_j].
\end{equation}

The cartesian orbitals with the spinor are labeled as $\ket{n_z,n_x,n_y,m_s}$, which are the single particle states that have been used in the Elliott SU(3) symmetry \cite{Elliott1,Elliott2}. From Eq. (\ref{R}) it follows that
\begin{equation}
[n_zn_xn_ym_s]=R\cdot [nlm_lm_s].
\end{equation}
Finally a transformation among the  Elliott states and the shell model states in matrix representation is \\ achieved
\begin{equation}\label{RS}
[n_zn_xn_ym_s]=R\cdot C\cdot [nljm_j].
\end{equation}
The inverse transformation is
\begin{equation}\label{SR}
[nljm_j]=C^{-1}\cdot R^\dagger [n_zn_xn_ym_s],
\end{equation}
where $R^\dagger$ is the conjugate transpose of $R$ and $C^{-1}$ is the inverse of $C$. 
The results of the transformations for the $\mathcal{N}=1-3$ shells are presented in Tables \ref{N1}-\ref{N3in} and for the $\mathcal{N}=1,2$ in Ref. \cite{SDANCA}. Results for higher shells, as well as the Mathematica code, by which they have been produced, are available from the first author upon request. 

The transformation presented in this section is consistent with the expansion coefficients $c(\lambda\mu K, L)$ of Ref. \cite{Elliott2}. For instance the orbital $\ket{n_z,n_x,n_y,m_s}$ =$\ket{2,0,0,{1\over 2}}$ corresponds to a $U(3)$ irrep $[f_1,f_2,f_3]=$$[n_z, n_x, n_y]$$=[2,0,0]$ and to the Elliott SU(3) quantum numbers $(\lambda,\mu)$ $=(f_1-f_2,f_2-f_3)=(2,0)$ \cite{Elliott1}. From Table \ref{N2} becomes evident, that this Elliott orbital expands with probability $\abs{-{1/\sqrt{3}}}^2={1/3}$ into an $s$ orbital and with probability $\abs{-{2/\sqrt{15}}}^2+\abs{\sqrt{2/ 5}}^2={2/ 3}$ into $d$ orbitals. The same expansion is presented in Table 3 of Ref. \cite{Elliott2} for the $(\lambda,\mu)=(2,0)$ irrep.

\section{The de Shalit--Goldhaber pairs}\label{deShalitG}

Sometimes  it is assumed, that the proton configuration is independent from that of the neutrons and thus one can treat protons and neutrons separately. However, studies of the proton-neutron interaction \cite{deShalit,FP1,FP2,Federman,Casten,Castenbook} have proved, that this interaction is so strong, that affects the $\beta$ decay transitions, the overall nuclear deformation and spectrum and gives rise to unexpected occupancies of the single particle nuclear states. 

De Shalit and Goldhaber \cite{deShalit} studied the $\beta$ transition probabilities and revealed, that the neutrons interact with the protons, when they occupy specific shell model orbitals. Their result, which is summarized at Figs. (5)-(8) of Ref. \cite{deShalit}, is, that the interaction of the neutrons of the $1i^{13/2},1h^{11/2}$ orbitals with the protons of the $1h^{11/2},1g^{9/2}$ orbitals respectively stabilizes the nucleus and, as a consequence, longer lifetimes in the $\beta$ transitions appear. The de Shalit--Goldhaber rule, that has been used in Refs. \cite{FP1,FP2,Federman}, is, that the valence neutrons bearing the highest possible $j_{max}=\mathcal{N}+{1\over 2}$  


\begin{table*}[htb]

\caption{The transformation matrix $R\cdot C$ for $\mathcal{N}=1$. The first line consists of the shell model orbitals, while the first column consists of the Elliott orbitals $\ket{n_z,n_x,n_y,m_s}$. These orbitals are used in the harmonic oscillator shell 2-8 ($p$ shell), or in the proxy-SU(3) shell 6-12 after the replacement of the intruder orbitals with their de Shalit--Goldhaber partners. See Section 3 for further discussion.}\label{N1}

\begin{tabular}{c c c c c c c}

\noalign{\smallskip}\hline\noalign{\smallskip}

$\ket{n,l,j,m_j}$ & $\ket{1p^{1/2}_{-1/2}}$ & $\ket{1p^{1/2}_{1/2}}$ & $\ket{1p^{3/2}_{-3/2}}$ & $\ket{1p^{3/2}_{-1/2}}$ & $\ket{1p^{3/2}_{1/2}}$ & $\ket{1p^{3/2}_{3/2}}$  \\
$\ket{n_z,n_x,n_y,m_s}$ &&&&&\\ 

\noalign{\smallskip}\hline\noalign{\smallskip}

$\ket{0,0,1,-{1\over 2}}$ &0 & $\frac{i}{\sqrt{3}}$ & $\frac{i}{\sqrt{2}}$ & 0 & $\frac{i}{\sqrt{6}}$ & 0 \\
$\ket{0,0,1,{1\over 2}}$ & $-\frac{i}{\sqrt{3}}$ & 0 & 0 & $\frac{i}{\sqrt{6}}$ & 0 & $\frac{i}{\sqrt{2}}$ \\
$\ket{0,1,0,-{1\over 2}}$ & 0 & $-\frac{1}{\sqrt{3}}$ & $\frac{1}{\sqrt{2}}$ & 0 & $-\frac{1}{\sqrt{6}}$ & 0 \\
$\ket{0,1,0,{1\over 2}}$ & $-\frac{1}{\sqrt{3}}$ & 0 & 0 & $\frac{1}{\sqrt{6}}$ & 0 & $-\frac{1}{\sqrt{2}}$ \\
$ \ket{1,0,0,-{1\over 2}}$ & $\frac{1}{\sqrt{3}}$ & 0 & 0 & $\sqrt{\frac{2}{3}}$ & 0 & 0 \\
$\ket{1,0,0,{1\over 2}}$ & 0 & $-\frac{1}{\sqrt{3}}$ & 0 & 0 & $\sqrt{\frac{2}{3}}$ & 0 \\

\noalign{\smallskip}\hline

\end{tabular}
\end{table*}


\begin{table*}[htb]

\caption{The inverse transformation matrix $C^{-1}\cdot R^\dagger$ for $\mathcal{N}=1$.}\label{N1in}

\begin{tabular}{c c c c c c c}
\noalign{\smallskip}\hline\noalign{\smallskip}\\

 $\ket{n_z,n_x,n_y,m_s}$ & $\ket{0,0,1,-{1\over 2}}$ &$\ket{0,0,1,{1\over 2}}$ &$\ket{0,1,0,-{1\over 2}}$&$\ket{0,1,0,{1\over 2}}$  & $\ket{1,0,0,-{1\over 2}}$ &$\ket{1,0,0,{1\over 2}}$ \\
 $\ket{n,l,j,m_j}$&&&&&\\ 

\noalign{\smallskip}\hline\noalign{\smallskip}

$\ket{1p^{1/2}_{-1/2}}$ & 0 & $\frac{i}{\sqrt{3}}$ & 0 & $-\frac{1}{\sqrt{3}}$ & $\frac{1}{\sqrt{3}}$ & 0 \\
$\ket{ 1p^{1/2}_{1/2}}$ & $-\frac{i}{\sqrt{3}}$ & 0 & $-\frac{1}{\sqrt{3}}$ & 0 & 0 & $-\frac{1}{\sqrt{3}}$ \\
$\ket{1p^{3/2}_{-3/2}}$ & $-\frac{i}{\sqrt{2}}$ & 0 & $\frac{1}{\sqrt{2}}$ & 0 & 0 & 0 \\
$\ket{1p^{3/2}_{-1/2}}$ & 0 & $-\frac{i}{\sqrt{6}}$ & 0 & $\frac{1}{\sqrt{6}}$ & $\sqrt{\frac{2}{3}}$ & 0 \\
$\ket{1p^{3/2}_{1/2}}$ & $-\frac{i}{\sqrt{6}}$ & 0 & $-\frac{1}{\sqrt{6}}$ & 0 & 0 & $\sqrt{\frac{2}{3}}$ \\
$\ket{1p^{3/2}_{3/2}}$ & 0 & $-\frac{i}{\sqrt{2}}$ & 0 & $-\frac{1}{\sqrt{2}}$ & 0 & 0 \\

\noalign{\smallskip}\hline

\end{tabular}
\end{table*}

\newpage


\begin{table*}[htb]
\caption{The same as Table \ref{N1}, but for $\mathcal{N}=2$, related the harmonic oscillator shell 8-20 ($sd$ shell), or to the proxy-SU(3) shell 14-26. }\label{N2}
\begin{tabular}{ccccccccccccc}

\noalign{\smallskip}\hline\noalign{\smallskip}\\

$\ket{n,l,j,m_j}$ & $\ket{2s^{1/2}_{-1/2}}$ & $\ket{2s^{1/2}_{1/2}}$ & $\ket{1d^{3/2}_{-3/2}}$ & $\ket{1d^{3/2}_{-1/2}}$ &  $\ket{1d^{3/2}_{1/2}}$ & $\ket{1d^{3/2}_{3/2}}$ & $\ket{1d^{5/2}_{-5/2}}$ & 
$\ket{1d^{5/2}_{-3/2}}$ & $\ket{1d^{5/2}_{-1/2}}$ & $\ket{1d^{5/2}_{1/2}}$ & $\ket{1d^{5/2}_{3/2}}$ & $\ket{1d^{5/2}_{5/2}}$ \\
$\ket{n_z,n_x,n_y,m_s}$ &&&&&\\ 

\noalign{\smallskip}\hline\noalign{\smallskip}

$\ket{0,0,2,-{1\over 2}}$& $-\frac{1}{\sqrt{3}}$ & 0 & 0 & $-\frac{1}{\sqrt{15}}$ & 0 & $-\frac{1}{\sqrt{5}}$ & $-\frac{1}{2}$ & 0 & $-\frac{1}{\sqrt{10}}$ & 0 & $-\frac{1}{2 \sqrt{5}}$ & 0 \\
$\ket{0,0,2,{1\over 2}}$& 0 & $-\frac{1}{\sqrt{3}}$ & $\frac{1}{\sqrt{5}}$ & 0 & $\frac{1}{\sqrt{15}}$ & 0 & 0 & $-\frac{1}{2 \sqrt{5}}$ & 0 & $-\frac{1}{\sqrt{10}}$ & 0 & $-\frac{1}{2}$\\
$\ket{0,1,1,-{1\over 2}}$& 0 & 0 & 0 & 0 & 0 & $-i \sqrt{\frac{2}{5}}$ & $\frac{i}{\sqrt{2}}$ & 0 & 0 & 0 & $-\frac{i}{\sqrt{10}}$ & 0 \\
$\ket{0,1,1,{1\over 2}}$& 0 & 0 & $-i \sqrt{\frac{2}{5}}$ & 0 & 0 & 0 & 0 & $\frac{i}{\sqrt{10}}$ & 0 & 0 & 0 & $-\frac{i}{\sqrt{2}}$ \\
$\ket{0,2,0,-{1\over 2}}$& $-\frac{1}{\sqrt{3}}$ & 0 & 0 & $-\frac{1}{\sqrt{15}}$ & 0 & $\frac{1}{\sqrt{5}}$ & $\frac{1}{2}$ & 0 & $-\frac{1}{\sqrt{10}}$ & 0 & $\frac{1}{2 \sqrt{5}}$ & 0 \\
$\ket{0,2,0,{1\over 2}}$& 0 & $-\frac{1}{\sqrt{3}}$ & $-\frac{1}{\sqrt{5}}$ & 0 & $\frac{1}{\sqrt{15}}$ & 0 & 0 & $\frac{1}{2 \sqrt{5}}$ & 0 & $-\frac{1}{\sqrt{10}}$ & 0 & $\frac{1}{2}$\\
$\ket{1,0,1,-{1\over 2}}$& 0 & 0 & $\frac{i}{\sqrt{10}}$ & 0 & $i \sqrt{\frac{3}{10}}$ & 0 & 0 & $i \sqrt{\frac{2}{5}}$ & 0 & $\frac{i}{\sqrt{5}}$ & 0 & 0 \\
$\ket{1,0,1,{1\over 2}}$& 0 & 0 & 0 & $-i \sqrt{\frac{3}{10}}$ & 0 & $-\frac{i}{\sqrt{10}}$ & 0 & 0 & $\frac{i}{\sqrt{5}}$ & 0 & $i \sqrt{\frac{2}{5}}$ & 0 \\
$\ket{1,1,0,-{1\over 2}}$& 0 & 0 & $\frac{1}{\sqrt{10}}$ & 0 & $-\sqrt{\frac{3}{10}}$ & 0 & 0 & $\sqrt{\frac{2}{5}}$ & 0 & $-\frac{1}{\sqrt{5}}$ & 0 & 0 \\
$\ket{1,1,0,{1\over 2}}$& 0 & 0 & 0 & $-\sqrt{\frac{3}{10}}$ & 0 & $\frac{1}{\sqrt{10}}$ & 0 & 0 & $\frac{1}{\sqrt{5}}$ & 0 & $-\sqrt{\frac{2}{5}}$ & 0 \\
$\ket{2,0,0,-{1\over 2}}$& $-\frac{1}{\sqrt{3}}$ & 0 & 0 & $\frac{2}{\sqrt{15}}$ & 0 & 0 & 0 & 0 & $\sqrt{\frac{2}{5}}$ & 0 & 0 & 0 \\
$\ket{2,0,0,{1\over 2}}$& 0 & $-\frac{1}{\sqrt{3}}$ & 0 & 0 & $-\frac{2}{\sqrt{15}}$ & 0 & 0 & 0 & 0 & $\sqrt{\frac{2}{5}}$ & 0 & 0 \\

\noalign{\smallskip}\hline

\end{tabular}
\end{table*}

within the shell, characterized by $\mathcal{N}$  quanta, overlap mostly with the valence protons bearing the highest $j'_{max}=(\mathcal{N}-1)+{1\over 2}$ within the previous shell, characterized by $\mathcal{N}-1$ quanta. Consequently the valence neutrons of the $1i^{13/2}, 1h^{11/2}, 1g^{9/2}, 1f^{7/2}, 1d^{5/2}$ attract the valence protons of the $ 1h^{11/2}, 1g^{9/2}, 1f^{7/2},$ $1d^{5/2}, 1p^{3/2}$ orbitals respectively. This mutual proton-neutron interaction has been attributed to a large spatial overlap among those orbitals. 

Similar results emerged \cite{Burcu2010} through the study of the quantity
\begin{eqnarray}
\abs{\delta V_{pn}(Z,N)}=\nonumber\\{1\over 4}[(B_{Z,N}-B_{Z,N-2})-(B_{Z-2,N}-B_{Z-2,N-2})]
\end{eqnarray}
for even-even nuclei, where $B$ is the binding energy. At Refs. \cite{Burcu2010,Sofia2013} the asymptotic Nilsson model notation \cite{Nilsson1,NR} has been used for the single particle states $K[\mathcal{N}n_z\Lambda]$. The conclusion was, that neutrons in the $K[\mathcal{N}n_z\Lambda]$ orbitals interact/overlap mostly with protons of the $K[\mathcal{N}-1,n_z-1,\Lambda]$ orbitals. Such proton-neutron orbitals differ in the Nilsson quantum numbers by\\ $\Delta K[\Delta\mathcal{N}\Delta n_z\Delta \Lambda]=0[110]$ and may be called $0[110]$ pairs. These $0[110]$ pairs appeared, to exhibit the maximum interaction, when both the proton and the neutron orbitals are the orbitals with the highest-$j$ allowed within their own shell\cite{Burcu2010,Sofia2013}, {\it i.e.}, when they originate from a de Shalit--Goldhaber pair of orbitals \cite{deShalit}. 

In this work we prove, that indeed the de Shalit--Goldhaber pairs differ by one quantum in the cartesian $z$-axis. The transformation of Section \ref{Trans} is used, to prove that
\begin{gather}
\mathcal{C} a_z \ket{1j^{15/2}_{m_j}}=\ket{1i^{13/2}_{m_j}},\label{a}\\
\mathcal{C} a_z \ket{1i^{13/2}_{m_j}}=\ket{1h^{11/2}_{m_j}},\label{b}\\
\mathcal{C} a_z \ket{1h^{11/2}_{m_j}}=\ket{1g^{9/2}_{m_j}},\label{c}\\
\mathcal{C} a_z\ket{1g^{9/2}_{m_j}}=\ket{1f^{7/2}_{m_j}},\label{d}\\
\mathcal{C} a_z\ket{1f^{7/2}_{m_j}}=\ket{1d^{5/2}_{m_j}},\label{e}\\
\mathcal{C} a_z\ket{1d^{5/2}_{m_j}}=\ket{1p^{3/2}_{m_j}},\label{f}
\end{gather}
where $\mathcal{C}$ are normalization constants (obtaining different values in each case), $a_z$ is the harmonic oscillator annihilation operator in the cartesian $z$-direction and $m_j$ has to be the same in the left and right side of each equation. The action of the $a_z$ operator is \cite{Cohen}
\begin{equation}\label{az}
a_z\ket{n_z}=\sqrt{n_z}\ket{n_z-1}.
\end{equation}

For clarity we present the equivalence of the $\mathcal{C}a_z\ket{d^{5/2}_{3/2}}$ to the $\ket{1p^{3/2}_{3/2}}$ orbital. The rest of the pairs are handled by a code. From Table \ref{N2in} the $\ket{1d^{5/2}_{3/2}}$ expands over the Elliott basis $\ket{n_z,n_x,n_y,m_s}$ as
\begin{eqnarray}\label{1d}
\ket{1d^{5/2}_{3/2}}=-{1\over 2\sqrt{5}}\ket{0,0,2,-{1\over 2}}+{i\over\sqrt{10}}\ket{0,1,1,-{1\over 2}}\nonumber\\+{1\over 2\sqrt{5}}\ket{0,2,0,-{1\over 2}}-i\sqrt{2\over 5}\ket{1,0,1,{1\over 2}}\nonumber\\-\sqrt{2\over 5}\ket{1,1,0,{1\over 2}}.
\end{eqnarray}
The action of the annihilation operator on this orbital is
\begin{equation}
a_z\ket{1d^{5/2}_{3/2}}=-i\sqrt{2\over 5}\ket{0,0,1,{1\over 2}}-\sqrt{2\over 5}\ket{0,1,0,{1\over 2}}.
\end{equation}
The normalization constant is $\mathcal{C}={\sqrt{5}\over 2}$. Therefore
\begin{equation}
\mathcal{C} a_z\ket{1d^{5/2}_{3/2}}=-{1\over\sqrt{2}}\ket{0,1,0,{1\over 2}}-{i\over\sqrt{2}}\ket{0,0,1,{1\over 2}},
\end{equation}
which from Table \ref{N1in} is equal to the $\ket{1p^{3/2}_{3/2}}$ orbital. The same equivalence stands for all the pairs of Eqs. (\ref{a})-(\ref{f}).

It is natural to wonder, if such an equivalence exists for other than the highest $j$ orbitals. For example, one can check, if the orbital $\ket{1d^{3/2}_{1/2}}$ differs only in the 
$z$-axis from the $\ket{1p^{1/2}_{1/2}}$ orbital. The same procedure yields
\begin{eqnarray}
\mathcal{C}a_z\ket{1d^{3/2}_{1/2}}\ne \ket{1p^{1/2}_{1/2}}.
\end{eqnarray}
Therefore the de Shalit--Goldhaber pairs are unique \hfill\break among the rest of the spin-orbit partners differing by \hfill\break $\ket{\Delta n,\Delta l,\Delta j,\Delta m_j} =\ket{0,1,1,0}$. The pairs of Eqs. (\ref{a})-(\ref{f}) differ by one quantum in the $z$-axis, thus they have similar structure in the $x$-$y$ plane. 

In the above considerations we have exploited the fact, that in both asymptotic deformed oscillator basis and cartesian Elliott basis the only change, inflicted by the proxy-SU(3) approximation, is the change of the quantum number $n_z$ (and therefore also of $\mathcal{N}$) by one unit. This similarity allows us, as a first step to translate in a simple way the proxy-SU(3) approximation from the asymptotic deformed oscillator  basis, in which it was defined in Ref. \cite{proxy1}, to the cartesian Elliott basis. At a second step the cartesian Elliott basis is connected in a non-trivial way to the spherical shell model basis, thus allowing us, to translate the proxy-SU(3) approximation into the spherical coordinates.      

\section{The importance of the $x$-$y$ plane}\label{xyplane}

Elliott has proved, that the rotational spectrum of the nucleus emerges from the SU(3) symmetry \cite{Elliott1}. This collective, rotational spectrum is derived by the occupied single particle orbitals in the cartesian coordinate system, with the nuclear wave function being characterized by the Elliott quantum numbers $(\lambda,\mu)$. 

The spatial, cartesian orbitals of a shell with $\mathcal{N}$ quanta are
\begin{eqnarray}\label{order}
\ket{n_z,n_x,n_y}:\ket{\mathcal{N},0,0}, \ket{\mathcal{N}-1,1,0}, \ket{\mathcal{N}-1,0,1},\nonumber\\ \ket{\mathcal{N}-2,2,0},\ket{\mathcal{N}-2,1,1},\ket{\mathcal{N}-2,0,2},...,\ket{0,0,\mathcal{N}}.\nonumber\\
\end{eqnarray}
The derivation of the many body U(3) irreps $[f_1,f_2,f_3]$ using the single particle states $\ket{n_z,n_x,n_y}$ as building blocks is discussed in detail in Refs. \cite{Draayer1989,Shukla, Rila}. Briefly the single particle orbitals are ordered in decreasing number of quanta $n_z$, because the low-lying nuclear properties are derived from the many particle, highest weight (hw) U(3) irrep 
$[f_1,f_2,f_3]$ \cite{Elliott2,proxy2} with $f_1,f_2,f_3$ being the summations for every valence nucleon:

\begin{gather}
f_1=\{\sum n_z\}_{max},\\
f_2=\max\{\sum n_x,\sum n_y\}, \mbox{ at } \{\sum n_z\}_{max},
\end{gather}

\begin{gather}
f_3=\min\{\sum n_x,\sum n_y\}, \mbox{ at } \{\sum n_z\}_{max}\\
 \mbox{ and }f_1\ge f_2\ge f_3.\nonumber
\end{gather}

For the low-lying nuclear properties the Elliott quantum numbers of the highest weight SU(3) irrep are \cite{Elliott2}
\begin{eqnarray}
\lambda=\{\sum n_z\}_{max}-\max\{\sum n_x,\sum n_y\},\label{l}\\
\mu=\abs{\sum n_x-\sum n_y},\mbox{ at } \{\sum n_z\}_{max}\label{xy}.
\end{eqnarray}
Consequently the quantum number $\mu$ depends solely on the distribution of quanta in the $x$-$y$ plane. For $\lambda\ge \mu$ the total angular momentum $J$ and its projection $K$ derive from the $(\lambda,\mu)$ \cite{Elliott4}:
\begin{eqnarray}
K_S=S,(S-1), \ldots, -S, \qquad K = K_S + K_L \geq 0, \\
 \mbox{where} \qquad K_L=\mu,\mu-2,\ldots,-\mu,\label{KL} \\
J=K,K+1, \ldots,(\lambda+\mu+S)\label{J},
\end{eqnarray}
where $S$, $K_S$ are the nuclear spin and its projection, with the exception that if $K_L=K_S=K=0$ then $J$ is even or odd according to $\lambda+S$.  

In simple words the quanta in the $x$-$y$ plane produce the quantum number $\mu$ from Eq. (\ref{xy}). Afterwards for $\lambda\ge \mu$ from Eq. (\ref{KL}) the quantum number $\mu$ derives all the $K_L$ bands within an irrep ,{\it i.e.} the ground state band with $K_L=0$, the $\gamma$ band with $K_L=2$, the $K=4$ band, etc. It becomes obvious from Eq. (\ref{J}), that the labeling $J$ of the states within a band depends on $K_L$, which depends on $\mu$, which depends on $\sum n_x, \sum n_y$. Thus the only quantum number left for the $\lambda$ to affect is the $J_{max}=\lambda+\mu+S$ value. The conclusion is, that the labeling of the bands and of the states within depends very much on the distribution of quanta in the $x$-$y$ plane and on $\mu$, while $\lambda$ affects the $J_{max}$ value within a band. Such $J_{max}$ states for medium mass and heavy nuclei, which are characterized by $\lambda\gg 1$ \cite{proxy2}, are never seen experimentally. As a result an accurate value of $\mu$ is of high importance, while an approximate value of $\lambda$ is not damaging the band structure. (It should not be overlooked, however, that $\lambda$ plays a crucial role in the predictions for the collective variable $\beta$, similar to the crucial role played by $\mu$ in the predictions for the collective variable $\gamma$ \cite{proxy2}.) 

The importance of the $x$-$y$ plane is also illustrated in the SU(3)$\rightarrow$SU(2)$\times$U(1) decomposition, that is presented in Section 2 of Ref. \cite{Elliott2}. The nucleus is a 3D object, which rotates in space, has ellipsoidal shape and possesses an SU(3) symmetry \cite{Shukla}. The ellipsoid is decomposed in the $x$-$y$ plane and in the $z$ symmetry axis. On the one hand the $x$-$y$ plane is a 2D object (an ellipsis), which is characterized by the SU(2) symmetry. The deformation of the ellipsis is measured by the $\mu$ quantum number of Eq. (\ref{xy}) and generates the projection of the total angular momentum $K$, which labels each band in the nuclear spectrum. On the other hand the $z$-axis possesses the U(1) symmetry and its elongation or compression in comparison with the $x$-$y$ plane is measured via Eq. (\ref{l}) and generates the cut-off of the total angular momentum. 

\section{Proxy-SU(3) symmetry in the spherical shell model basis}

The main idea behind the proxy-SU(3) symmetry in medium mass and heavy nuclei, introduced in \cite{proxy1}, is that one can restore the SU(3) symmetry within the spin-orbit shells among the magic numbers 28, 50, 82, 126 by replacing the intruder asymptotic deformed oscillator orbitals by their $\Delta K[\Delta \mathcal{N}\Delta n_z\Delta\Lambda]=0[110]$ counterparts. Such pairs of orbitals in the Nilsson notation differ only in the cartesian $z$-axis and are identical in the $x$-$y$ plane. 

Similarly in this work we introduce the proxy-SU(3) symmetry using the spherical shell model basis. In each of the above mentioned spin-orbit shells the replacement of the intruder orbitals $1g^{9/2}$, $1h^{11/2}$, $1i^{13/2}$ by the orbitals $1f^{7/2},$ $1g^{9/2},1h^{11/2}$ respectively with the same $m_j$ leads to the reconstruction of the SU(3) symmetry. The intruder orbitals with the highest $\abs{m_j}=j_{max}$ are totally annihilated by the $a_z$ operator, thus they do not have any $\ket{\Delta n,\Delta l,\Delta j,\Delta m_j}=\ket{0,1,1,0}$ partner. Consequently the spin-orbit like shells among magic numbers 6-14, 14-28, 28-50, 50-82, 82-126, 126-184 become 6-12, 14-26, 28-48, 50-80, 82-124, 126-182 shells after the replacement of the intruder orbitals. The replacements are presented in detail in Table \ref{shells}. 

Although the de Shalit--Goldhaber rule has been discovered for proton-neutron pairs, it is valid, that the proton-proton or neutron-neutron de Shalit--Goldhaber pairs of orbitals (Eqs. (\ref{a})-(\ref{f})) have similar structure in the $x$-$y$ plane.  Thus, as outlined in Section \ref{xyplane}, the replacement of these orbitals by one another is not damaging the Elliott quantum number $\mu$ in prolate nuclei. Consequently the $K$ rotational bands and the $J$ states within these bands are labeled accurately  within the proxy-SU(3) symmetry. The only quantum number affected by the proposed replacement is the $f_1=\sum n_z$ or $\lambda$ from Eq. (\ref{l}), which is relevant with the cut-off of the total angular momentum $J_{max}$ in each band, as pointed out in the previous section. But in medium mass and heavy nuclei the $J_{max}$ value is already too large, 
to be observed experimentally \cite{proxy1,proxy2}.

In summary, in this section we have exploited the fact, that the de Shalit--Goldhaber pairs have identical structure in the $x$-$y$ plane, differing only by one quantum along the 
$z$-axis. Taking advantage of the transformation between the cartesian Elliott basis and the spherical shell model basis, established in Sections 2 and 3, we were able to ``translate'' the  proxy-SU(3) substitution rule from the asymptotic deformed oscillator basis, in which it is $\Delta K[\Delta \mathcal{N}\Delta n_z\Delta\Lambda]=0[110]$, to the spherical shell model basis, where it is $\ket{\Delta n,\Delta l,\Delta j,\Delta m_j}=\ket{0,1,1,0}$.  It should be noticed, that while in the asymptotic deformed oscillator basis it is evident, that \textsl{the projections} of the orbital and the total angular momenta and the spin remain unchanged, in the spherical shell model basis becomes evident, that the eigenvalues of the orbital and the total angular momenta are changed by one unit. In other words, different bases offer different ``views'' of the same physical quantities, namely angular momenta and spin in the present case.

\section{Towards a  unitary transformation}

In the equations (\ref{a})-(\ref{f}) of Section \ref{deShalitG} we can express the normalization constant $\mathcal{C}$ in terms of the $a_z^\dagger$, $a_z$ operators. The normalization constant is the expectation value
\begin{eqnarray}
\mathcal{C}=\braket{n,l,j,m_j|a_z^\dagger a_z|n,l,j,m_j}^{-1/2}\nonumber\\=<a_z^\dagger a_z>^{-1/2}=<n_z>^{-1/2}.
\end{eqnarray}
Consequently the operator
\begin{equation}\label{T}
T=a_z<a_z^\dagger a_z>^{-1/2}
\end{equation}
acts on the intruder states, which possess $j_{max}=\mathcal{N}+1/2$ and $-(j_{max}-1) \le m_j\le (j_{max}-1)$, as
\begin{equation}\label{act}
T\ket{n,l,j_{max},m_j}=\ket{n,l-1,j_{max}-1,m_j}.
\end{equation}
It should be noted, that the above action is true only for the intruder states and that the intruder states with $m_j=\pm j_{max}$ are annihilated by $T$. A graphical representation of an example of the action of the operator $T$ is given in Fig. \ref{proxies}. 

For instance in the $1d^{5/2}_{3/2}$ orbital of Eq. (\ref{1d}) we have 
\begin{eqnarray}
<a_z^\dagger a_z>=\braket{1d^{5/2}_{3/2}|a_z^\dagger a_z|1d^{5/2}_{3/2}}=\nonumber\\
={2\over 5}\cdot 1+{2\over 5}\cdot 1={4\over 5}, 
\end{eqnarray}
and thus
\begin{equation}
\mathcal{C}={\sqrt{5}\over 2}.
\end{equation}
In this case
\begin{eqnarray}
T\ket{1d^{5/2}_{3/2}}=a_z\braket{1d^{5/2}_{3/2}|a_z^\dagger a_z|1d^{5/2}_{3/2}}^{-1/2}\ket{1d^{5/2}_{3/2}}=\nonumber\\
\mathcal{C}a_z \ket{1d^{5/2}_{3/2}}=\ket{1p^{3/2}_{3/2}}.
\end{eqnarray}

The operator $T$ resembles the unitary operator used in Ref. \cite{AnnArbor} in the case of the harmonic oscillator, reviewed in Appendix A. In the case of the 3D-HO without spin one can use the unitary operator
\begin{equation}\label{Uop}
U=a_z(a_z^\dagger a_z)^{-1/2},
\end{equation}
affecting only the $z$-direction and leaving the $x$, $y$ directions intact.

In relation to the prerequisites mentioned in Appendix A, in the present case we use the Hermitian operator
\begin{equation}
[\mathcal{O}(\mathbf{r},\mathbf{p},S)]^{-1/2}=(a_z^\dagger a_z)^{-1/2}.
\end{equation}
Since $\mathcal{O}$ is Hermitian, it is possible, to find a basis, in which it is diagonal. Furthermore, in this basis the $\mathcal{O}^{-1/2}$ makes sense, as we take the square root \cite{Halmos1,Halmos2,Bernau,Sebestyen} 
of the inverse terms in the diagonal, if the eigenvalues of $\mathcal{O}$ are positive. In the present case the operator $a_z^\dagger a_z$ is diagonal in the $\ket{n_z}$ basis and has eigenvalues \cite{Cohen}:
\begin{equation}
a_z^\dagger a_z\ket{n_z}=n_z\ket{n_z}.
\end{equation}
Therefore
\begin{equation}
(a_z^\dagger a_z)^{-1/2}=(n_z)^{-1/2}.
\end{equation}
Obviously the operator $a_z^\dagger a_z$ is not diagonal in the shell model states $\ket{n,l,j,m_j}$. This is why we are using the expectation value of this operator in Eq. (\ref{T}).

\section{Unitary transformation for the proxy-SU(3) scheme}

The question is now created, about what would have been an appropriate unitary transformation, performing the proxy-SU(3) change within the spherical shell model. From Section 5  we know, that in the notation 
$\ket{n,l,j,m_j}$ the appropriate change is $\ket{0,1,1,0}$. Qualitatively we would expect a transformation affecting only $\mathcal{N}$ and $l$, since in spherical coordinates one has $\mathcal{N}= 2n +l$, where $n$ is a non-negative integer. Therefore the only existing possibility in this case, appears to be, that the loss of one unit of $\mathcal{N}$ corresponds to the loss of one unit of $l$. However, in quantum mechanics we are familiar with the ladder operators $l_+$, $l_-$, which correspondingly raise or lower the projection $m_l$ of $l$ by one unit \cite{Cohen}, but we are not familiar with any simple ladder operators changing $l$ or $j$ by one unit. Therefore the task of constructing a unitary transformation in this case seems impossible.

However the mapping between the cartesian Elliott states and the spherical shell model states,  worked out in Section 2, offers a way out. Eqs. (\ref{a})-(\ref{f}) show, that exactly for these orbitals, which become abnormal parity orbitals because of the spin-orbit interaction, the annihilation operator $a_z$ can be used. Indeed, Eqs. (\ref{a})-(\ref{f}) show, that the $a_z$ operator reduces $l$ and $j$ by one
unit, without changing the projection $m_j$. In other words, one can perform the unitary transformation in the cartesian Elliott basis, using the unitary transformation corresponding to the $a_z$ operator in the cartesian Elliott basis, and then translate the results into the language of the spherical shell model. Simply speaking the operators $T$ and $U$ of Section 6 are the same operator acting on different bases: the operator $T$ is acting on the shell model states $\ket{n,l,j,m_j}$, while the operator $U$ is acting on the Elliott states $\ket{n_z,n_x,n_y,m_s}$. 

The operator $T$ preserves the inner product of any two intruder states, possessing $j=j_{max}=\mathcal{N}+1/2$ and $-(j_{max}-1)\le m_j\le (j_{max}-1)$. Indeed the inner product is
\begin{equation}
\braket{n,l,j_{max},m'_j|n,l,j_{max},m_j}=\delta_{m'_j,m_j},
\end{equation}
while the inner product of the states after the action of $T$ is
\begin{equation}
\braket{Tn,l,j_{max},m'_j|Tn,l,j_{max},m_j}, 
\end{equation}
where the shorthand notation 
\begin{equation}
T\ket{n,l,j_{max},m_j}= \ket{Tn,l,j_{max},m_j}
\end{equation}
has been used. 
The substitution of Eq. (\ref{act}) in the above results to
\begin{eqnarray}\label{product}
\braket{Tn,l,j_{max},m'_j|Tn,l,j_{max},m_j}=\nonumber\\\braket{n,l-1,j_{max}-1,m'_j|n,l-1,j_{max}-1,m_j}=\delta_{m'_j,m_j}.\nonumber\\
\end{eqnarray}
Consequently, since the inner product is preserved, the operator $T$ is unitary, when acting on the intruder states with $j_{max}$ and $-(j_{max}-1)\le m_j\le j_{max}-1$:
\begin{eqnarray}
\braket{Tn,l,j_{max},m'_j|Tn,l,j_{max},m_j}=\nonumber\\\braket{n,l,j_{max},m'_j|n,l,j_{max},m_j}.
\end{eqnarray}
The matrix elements of the operator $T^\dagger T$ in the basis of the intruder shell model orbitals $\ket{n,l,j_{max},m_j}$ with $j_{max}=\mathcal{N}+{1\over 2}$ and $-(j_{max}-1)\le m_j\le j_{max}-1$ are calculated via Eq. (\ref{product})
\begin{eqnarray}
\braket{n,l,j_{max},m_j'|T^\dagger T|n,l,j_{max},m_j}=\nonumber\\
\braket{T n,l,j_{max},m_j'|Tn,l,j_{max},m_j}=
\delta_{m_j',m_j}.
\end{eqnarray}
Thus the operator $T^\dagger T$ is the square identity matrix 
\begin{equation}
T^\dagger T=I
\end{equation}
with dimension $(2j_{max}-1)\times (2j_{max}-1)$. If the two annihilated intruder orbitals with $m_j=\pm j_{max}$ are included in the matrix representation, for which
\begin{equation}
T\ket{n,l,j_{max},\pm j_{max}}=0,
\end{equation}
then the operator $T^\dagger T$ is no longer the identity matrix $I$ and the mapping from the intruders to the proxies is represented by a rectangular not by a square operator, since two of the intruders (those with $m_j=\pm j_{max}$) have no images in the proxy space. A generalization of the above consideration is, that {\it the transformation $T$ is unitary, when acting on the intruder states with $j_{max}=\mathcal{N}+{1\over 2}$ and $-(j_{max}-1)\le m_j\le j_{max}-1$}.

Therefore one can apply the unitary transformation on all except from the $\abs{m_j}=j_{max}$ the abnormal parity orbitals, thus forming a full $U(\Omega)$ HO shell, composed by the unitary images of the abnormal parity orbitals and the unaffected normal parity orbitals. As we have shown above, the proxy-SU(3) substitution rule, which is $\Delta K[\Delta \mathcal{N}\Delta n_z\Delta\Lambda]=0[110]$ in the asymptotic deformed oscillator basis, is translated into the rule $\ket{\Delta n,\Delta l,\Delta j,\Delta m_j}$ $=\ket{0,1,1,0}$ in the spherical shell model basis. This correspondence is based on the fact, that the proxy-SU(3) approximation is only affecting one quantum in the $z$-axis, leaving the $x$-$y$ plane unchanged. As a consequence, the changes inflicted by the proxy-SU(3) approximation in the cartesian Elliott basis and in the asymptotic deformed oscillator basis can be simply expressed in both cases by using the $a_z$ annihilation operator and the unitary $T$ operator of Eq. (\ref{T}) related to it.

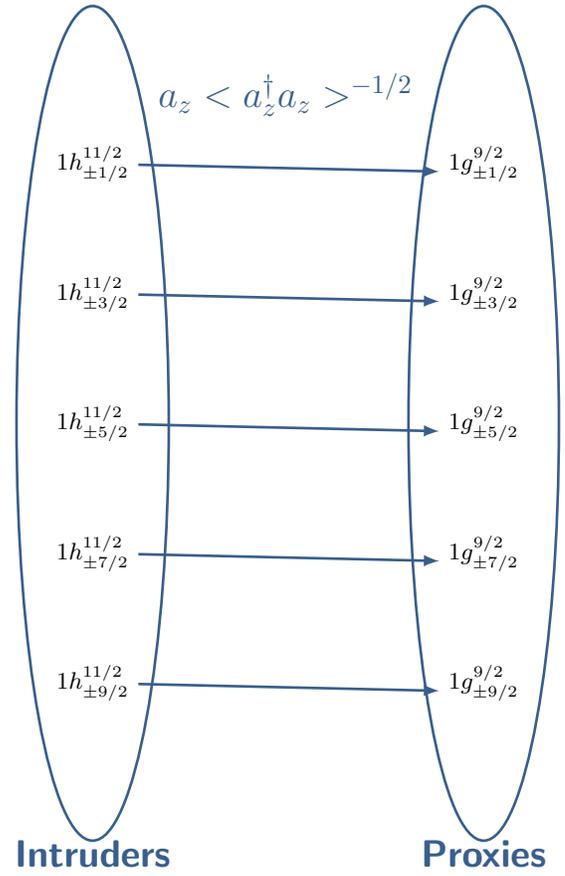
\begin{figure}
\begin{tikzpicture}[line width=1pt,>=latex]
\sffamily
\node (a1) {$1h^{11/2}_{\pm 1/2}$};
\node[below=of a1] (a2) {$1h^{11/2}_{\pm 3/2}$};
\node[below=of a2] (a3) {$1h^{11/2}_{\pm 5/2}$};
\node[below=of a3] (a4) {$1h^{11/2}_{\pm 7/2}$};
\node[below=of a4] (a5) {$1h^{11/2}_{\pm 9/2}$};

\node[above right=0.2cm of a1] (b1) {\color{myblue}\Large $ a_z<a_z^\dagger a_z>^{-1/2}$};
\node[right=4cm of a1] (b1) {$1g^{9/2}_{\pm 1/2}$};
\node[below= of b1] (b2) {$1g^{9/2}_{\pm 3/2}$};
\node[below=of b2] (b3) {$1g^{9/2}_{\pm 5/2}$};
\node[below=of b3] (b4) {$1g^{9/2}_{\pm 7/2}$};
\node[below=of b4] (b5) {$1g^{9/2}_{\pm 9/2}$};

\node[shape=ellipse,draw=myblue,minimum size=2cm,fit={(a1) (a5)}] {};
\node[shape=ellipse,draw=myblue,minimum size=2cm,fit={(b1) (b5)}] {};

\node[below=1.6cm of a5,font=\color{myblue}\Large\bfseries] {Intruders};
\node[below=1.6cm of b5,font=\color{myblue}\Large\bfseries] {Proxies};

\draw[->,myblue] (a1) -- (b1.190);
\draw[->,myblue] (a2) -- (b2.190);
\draw[->,myblue] (a3) -- (b3.190);
\draw[->,myblue] (a4) -- (b4.190);
\draw[->,myblue] (a5) -- (b5.190);
\end{tikzpicture}
\caption{In the 50-82 shell the intruders $1h^{11/2}_{m_j}$ (except for the $1h^{11/2}_{\pm 11/2}$) are mapped onto the $1g^{9/2}_{m_j}$ through the operator  
$T=a_z<a_z^\dagger a_z>^{-1/2}$. The inner product of any two intruders with $j=j_{max}$ and $-(j_{max}-1)\le m_j\le (j_{max}-1)$ remains the same before and after the transformation, {\it i.e.},
$\braket{n,l,j_{max},m'_j|n,l,j_{max},m_j}$= $\braket{Tn,l,j_{max},m'_j|Tn,l,j_{max}m_j}$, which equals to $\braket{n,l-1,j-1,m'_j|n,l-1,j-1,m_j}$= $\delta_{m'_j,m_j}$. Thus the operator $T$ is unitary, when acting on the intruder orbitals of each spin-orbit like shell with $j=j_{max}$ and $-(j_{max}-1)\le m_j\le (j_{max}-1)$. The intruders with $m_j=\pm j_{max}$ are being totally annihilated and so they are left out of the mapping. }\label{proxies}
\end{figure}

As an example, the 50-82 nuclear shell consists of the $3s^{1/2}$, $2d^{3/2}$, $2d^{5/2}$, $1g^{7/2}$, and $1h^{11/2}$ orbitals. In comparison to the $sdg$ 3D-HO shell, it has lost the $1g^{9/2}$ orbital, which is pushed down by the spin-orbit interaction into the shell below, and has gained the $1h^{11/2}$ orbital, which has come down, again because of the spin-orbit interaction, from the shell above. Through the unitary transformation under discussion, the orbital  $1h^{11/2}$ is mapped onto the proxy-orbital $1g^{9/2}$, forming together with the $3s^{1/2}$, $2d^{3/2}$, $2d^{5/2}$, $1g^{7/2}$ orbitals a complete $sdg$ shell. In particular, the $1h^{11/2}_{\pm 1/2}$, $1h^{11/2}_{\pm 3/2}$, $1h^{11/2}_{\pm 5/2}$, $1h^{11/2}_{\pm 7/2}$, $1h^{11/2}_{\pm 9/2}$ levels are transformed into the $1g^{9/2}_{\pm 1/2}$, $1g^{9/2}_{\pm 3/2}$, $1g^{9/2}_{\pm 5/2}$, $1g^{\pm 9/2}_{7/2}$, $1g^{9/2}_{\pm 9/2}$ levels as shown in Fig. \ref{proxies}, while the levels $1h^{11/2}_{\pm 11/2}$ are left out of the transformation.

It should be pointed out, that the $+1$ term in the Hamiltonian, appearing in Eqs. (\ref{HO}) and (\ref{HO2}), is of utmost importance within the proxy-SU(3) approximation scheme, since this term guarantees, that the proxies of the abnormal parity orbitals do not slip into the shell below, but remain within the spin-orbit shell, in which they are already located, thus completing together with the normal parity orbitals a full HO shell possessing the relevant $U(\Omega)$ symmetry. In Ref. \cite{proxy1} this constant term has been added ``by hand'', using physical arguments. In the present work it is proved, that this term arises from the unitary transformation between the abnormal parity levels and their proxies. 

The $+1$ term in the Hamiltonians of Eqs.  (\ref{HO}) and (\ref{HO2}), in the Appendix A and B respectively, guarantees, that the proxy orbital $1g^{9/2}$ stays within the 50-82 shell, in which the abnormal parity orbital $1h^{11/2}$ existed. Therefore all orbitals 3s$^{1/2}$, 2d$^{3/2}$, $2d^{5/2}$, $1g^{7/2}$, 1g$^{9/2}$ are in the same shell, forming a U(15) algebra having an SU(3) subalgebra. The $+1$ term in Eqs. (\ref{HO}) and (\ref{HO2}) also clarifies  the question, about the form of the Hamiltonian to be used within the proxy-SU(3) scheme: It is the usual Hamiltonian for the shell under discussion, using the parameters of the normal parity orbitals with $\mathcal{N}$ quanta. In Ref. \cite{proxy1}, for example, in the 50-82 shell the abnormal parity orbital $1h^{11/2}$, which has come down from the 82-126 shell having been calculated from the Nilsson Hamiltonian with the parameters of the orbitals with $\mathcal{N}+1=5$, is transformed into a $1g^{9/2}$ orbital, which, from this point on, is treated using the parameters of the orbitals with $\mathcal{N}=4$. Obviously this is an approximation, since in Eqs. (\ref{HO}) and (\ref{HO2}) the Hamiltonians $H_0$ and $H'_0$, connected by the unitary transformation, possess the same parameter set. Furthermore, one can see from Table I of Ref. \cite{proxy1}, that the change of the parameter values, when moving from one shell to the next one, is not dramatic. However, this approximation does have a non-negligible effect on the relevant Nilsson diagrams, as seen for example in Fig. 2 of Ref. \cite{proxy1}, namely that the proxies of the abnormal parity levels do not converge to a single point at zero deformation.      

\section{Comparison to the unitary transformation for the pseudo-SU(3) scheme within the spherical Nilsson model} 

The problem of finding an approximate SU(3) symmetry in a subspace of the total Hilbert space was addressed earlier within the pseudo-SU(3) scheme
\cite{Adler,Shimizu,pseudo1,pseudo2,DW1,DW2,Harwood,Ginocchio1,Ginocchio2}. There is a significant similarity between the pseudo-SU(3) and proxy-SU(3) approaches, although the approximations they apply are different.  In the pseudo-SU(3) scheme all the intruder levels are skipped, and the rest of the states correspond to a HO shell of one oscillator quantum less. Therefore, usually SU(3) irreps of smaller dimensions are relevant within the pseudo-SU(3) scheme. A considerable number of nucleons are out of the SU(3) subspace, and they are usually treated in terms of the seniority formalism. In the proxy-SU(3) scheme only two single orbitals are omitted (those with $\abs{m_j}=j_{max}$), so usually SU(3) irreps of higher dimensions are relevant, while no other sector is coupled to the HO obtained within the proxy-SU(3) scheme.  Due to the different approximations used, the performance and the applicability of the two models can be different.  Nevertheless, it may also very well happen, that their performance is similar for some phenomena. The decisive factor seems to be the relevant subset of the single-particle orbits, which may be similar or different for the two cases.  A recent application to quartet spectra including also major shell excitations showed, that the physics of the two approaches is very similar (in the case investigated), in spite of the different SU(3) quantum numbers occurring in each of them \cite{Cseh}. In addition, recent calculations of the collective variables $\beta$ and $\gamma$ within the two different approximation schemes have been found, to lead to compatible results \cite{EPJST}. 

The unitary transformation applicable in the case of the pseudo-SU(3) symmetry, described in Appendix B, has been applied for clarification purposes to the spherical Nilsson Hamiltonian \cite{AnnArbor,Quesne}
\begin{equation}
{H\over \hbar \omega} = H_0 - 2k {\bf L}\cdot {\bf S} -k \nu \mathbf{L}^2, 
\end{equation}
where $H_0$ is the 3D-HO Hamiltonian defined in Appendix B, while $k$ and $\nu$ are free parameters, adjusted so that the experimental single-particle energies are well reproduced. In this case the unitary operator is obtained from $({\bf a} \cdot {\bf S})$, as given in Eq. (\ref{US}). The Hamiltonian after the unitary transformation reads  \cite{AnnArbor,Quesne}
\begin{equation}
{H'\over \hbar \omega} = H_0+1 - 2k(2\nu-1) {\bf L}\cdot {\bf S} -k\nu \mathbf{L}^2 -2k(\nu-1). 
\end{equation}
Denoting by $l(l+1)$, $s(s+1)$, $j(j+1)$ the eigenvalues of the operators ${\bf L}^2$, ${\bf S}^2$, 
${\bf J}^2$, we notice in this case, that the unitary transformation keeps invariant the total angular momentum $j$, while $\mathcal{N}$, $l$, $s$ are mapped onto their pseudo counterparts $\mathcal{\tilde N}$, 
$\tilde l$ and $\tilde s$, so that the correct $\tilde j= j $ is obtained. In the case of $k=\nu=0$ the spherical Nilsson results are reduced to the usual 3D-HO results.

As an example, the 50-82 nuclear shell consists of the $3s^{1/2}$, $2d^{3/2}$, $2d^{5/2}$, $1g^{7/2}$, and $1h^{11/2}$ orbitals. In comparison to the $sdg$ 3D-HO shell, it has lost the $1g^{9/2}$ orbital, which is pushed down by the spin-orbit interaction into the shell below, and has gained the $1h^{11/2}$ orbital, which has come down, again because of the spin-orbit interaction, from the shell above. Through the unitary transformation under discussion, the orbitals $3s^{1/2}$, $2d^{3/2}$, 2d$^{5/2}$, $1g^{7/2}$, are mapped onto the pseudo-orbitals $2p^{1/2}$, $2p^{3/2}$, $1f^{5/2}$, $1f^{7/2}$, forming a complete $pf$ shell. During this mapping, $j$ remains invariant, while $\mathcal{N}$, $l$, $s$ are appropriately changed in order to guarantee the invariance of $j$.  

\section{Discussion}

The main results, obtained in this work, are the following ones.

a) Using the connection between the cartesian  Elliott basis and the spherical shell model basis it is shown, that the proxy-SU(3) approximation within the spherical shell model framework corresponds to the replacement of the abnormal parity orbitals in a shell by their de Shalit--Goldhaber counterparts.

b)It is shown, that the replacement of the abnormal parity orbitals within a shell by their de Shalit--Goldhaber counterparts is equivalent to a unitary transformation, affecting only the $z$-axis, while leaving the $x$-$y$ plane and the $z$-projections of the angular momenta and the spin intact.

Furthermore, the following points should be emphasized.

a)The present work involves the pure SU(3) symmetry of the usual 3D-HO, appearing in the Elliott model, thus bypassing the deformed SU(3) symmetry connected to the Nilsson model and the mathematical complications, related to it \cite{Asherova,Lenis}.  
   
b)The present work paves the way for shell model calculations involving the proxy-SU(3) approximation, since the appropriate replacements to be used within the spherical shell model, have been determined. As a first step, one can check numerically the accuracy of the proxy-SU(3) approximation by performing a calculation of accessible size twice, without and with the proxy-SU(3) substitution, and comparing the results. At a next stage, one can try to cut down the size of large shell model calculations by taking advantage of the SU(3) properties. 

c)The difference between a classification scheme and a dynamical symmetry is clarified. Proxy-SU(3) is a classification scheme valid through the entire nuclear shells, in the same way in which the Elliott SU(3) is valid throughout the whole $sd$ shell. The proxy-SU(3) dynamical symmetry will be needed, in order to calculate spectra and electromagnetic transition probabilities in deformed nuclei, after choosing an appropriate Hamiltonian involving three- and/or four-body terms \cite{DW1,DW2}. Work in this direction is in progress. 

d)In corroboration of the previous point one can recall, that in the first Elliott paper \cite{Elliott1} an alternative classification scheme for the $sd$ shell is given in terms of the O(6) subalgebra. In other words, different classification schemes can be used for the entire $sd$ shell irrespectively of the dynamical symmetries, possibly describing the spectra and electromagnetic transition rates of various nuclei within this shell. One should notice, however, that the O(6) classification scheme is not possible in all higher shells.  

\section*{Acknowledgements} 

Financial support by the Greek State Scholarships Foundation (IKY) and the European Union within the MIS 5033021 action, by the Bulgarian National Science Fund (BNSF) under Contract No. KP-06-N28/6 and by National Research, Development and Innovation Fund of Hungary, under the K18 funding scheme with Project No. K 128729 is gratefully acknowledged.

\section*{Appendix A: Prerequisites for the unitarity of the operators}

Within the pseudo-SU(3) approach it is well known, that there is a unitary transformation connecting the normal parity orbitals in a nuclear shell to their pseudo-SU(3) counterparts. A pedagogical description of this unitary transformation is given in \cite{AnnArbor}, applied to the case of the one-dimensional harmonic oscillator (1D-HO) (Section 4 of \cite{AnnArbor}). Subsequently, the unitary transformation for the three-dimensional harmonic oscillator (3D-HO) from spherical coordinates to the pseudo-SU(3) basis is given in Section 5 of \cite{AnnArbor}, while the unitary transformation, appropriate for the spherical Nilsson Hamiltonian (with deformation $\epsilon=0$), which includes a spin-orbit interaction term and an $\mathbf{l}^2$ term, where $\mathbf{l}$ is orbital angular momentum, is given in Section 6 of \cite{AnnArbor}. A concise version of this work is given in \cite{Quesne}. A generalization of this unitary transformation to the case of the deformed Nilsson Hamiltonian has been given in \cite{Hess}.

Let us consider an operator $G(\mathbf{r}, \mathbf{p}, \mathbf{S})$, where $\mathbf{r}$, $\mathbf{p}$, $\mathbf{S}$ are the position, momentum, and spin respectively. The operator $G$ will not be in general unitary, but it can be made unitary through the use of the Hermitian operator \cite{AnnArbor}
\begin{equation}
\mathcal{O}(\mathbf{r}, \mathbf{p}, \mathbf{S}) = G^\dagger(\mathbf{r}, \mathbf{p}, \mathbf{S})  G(\mathbf{r}, \mathbf{p}, \mathbf{S}),
\end{equation}
by taking 
\begin{equation}
U = G(\mathbf{r}, \mathbf{p}, \mathbf{S}) [\mathcal{O} (\mathbf{r}, \mathbf{p}, \mathbf{S})]^{-1/2}. 
\end{equation}
Since $\mathcal{O}$ is a Hermitian operator, it is possible, to find a basis, in which this operator is diagonal. In this basis the operator $[\mathcal{O} (\mathbf{r}, \mathbf{p}, \mathbf{S})]^{-1/2}$ can be obtained by taking the square roots of  the inverse terms in the diagonal, provided that the eigenvalues of $\mathcal{O}$ are positive. Square roots of positive operators have been considered repeatedly in the mathematical literature \cite{Halmos1,Halmos2,Bernau,Sebestyen}. 

In the case of the original 1D-HO Hamiltonian one has \cite{Cohen}:
\begin{equation}
H_0 = a^\dagger a, \quad  a^\dagger= {1\over \sqrt{2}}(x-ip), \quad  a= {1\over \sqrt{2}}(x+ip),
\end{equation}
where $x$ and $p$ are the coordinate and the corresponding momentum, 
while $a^\dagger$ and $a$ denote the usual creation and annihilation operators. The unitary operator is obtained from the usual annihilation operator $a$ and has the form \cite{AnnArbor}
\begin{equation} 
U =  a (a^\dagger a)^{-1/2}.
\end{equation}
It is clear, that the above mentioned prerequisites are fulfilled, since the operator $a^\dagger a$ in the basis considered is diagonal and has positive eigenvalues, so that the square roots of their inverses can be calculated. 
The Hamiltonian resulting from the unitary transformation is \cite{AnnArbor}
\begin{equation}\label{HO}
H_0' = a^\dagger a  +1.  
\end{equation}
This implies, that the levels characterized by $\mathcal{N}$ quanta in the old Hamiltonian carry $\mathcal{N}-1$ quanta in the new one, sharing the same energy, while the ground state of the old Hamiltonian has no counterpart. In other words, in the new Hamiltonian $\mathcal{N}$ is replaced by $\mathcal{N}-1$, but the loss of energy is counterbalanced by the $+1$ term appearing in Eq. (\ref{HO}). This $+1$ term plays a major role in the proxy-SU(3) approximation, as discussed in Section 7. 

In the above the spectra of the old and the new Hamiltonian are the same, as they should be for a unitary transformation. It should also be noticed, that we have the complete set of states of the harmonic oscillator both before and after the unitary transformation, {\it i.e.}, the ground state is included in both cases. Therefore the U(1) symmetry characterizing the 1D-HO before the unitary transformation is conserved after the unitary transformation. 

\section*{Appendix B: Unitary transformations in the \hfill\break pseudo-SU(3) scheme}\label{pseudo}

For the sake of comparison, the original 3D-HO Hamiltonian for a particle with spin 1/2 is: 
\begin{equation}
H_0 = \mathbf{a^\dagger } \cdot \mathbf{a}, \quad  {\bf{a^\dagger}}= {1\over \sqrt{2}}(\mathbf{r}-i \mathbf{p}), \quad  {\bf a}= {1\over \sqrt{2}}({\bf r}+i{\bf p}),
\end{equation}
where ${\bf r}$ and ${\bf p}$ are the coordinates and the corresponding momenta, while $\bf{a^\dagger}$ and ${\bf a}$ denote the usual creation and annihilation operators. 
The unitary operator is obtained from $({\bf a} \cdot {\bf S})$, where ${\bf S}$ is the spin operator, and reads \cite{AnnArbor}
\begin{equation}\label{US}
U =2(\mathbf{a} \cdot \mathbf{S}) 
(\mathbf{a^\dagger} \cdot \mathbf{a} -2 \mathbf{L} \cdot \mathbf{S}) ^{-1/2}.
\end{equation}
The Hamiltonian resulting from the unitary transformation is \cite{AnnArbor}:
\begin{equation}\label{HO2}
H_0' = {\bf a^\dagger}\cdot  {\bf a} +1.  
\end{equation}
This again implies, that the levels characterized by $\mathcal{N}$ quanta in the old Hamiltonian carry $\mathcal{N}-1$ quanta in the new one, sharing the same energy, while the ground state of the old Hamiltonian has no counterpart.  As in the case of the 1D-HO, the loss of one quantum of excitation energy by the fact that $\mathcal{N}$ of the original Hamiltonian is mapped onto $\mathcal{N}-1$ in the final Hamiltonian is counterbalanced by the $+1$ term in Eq. (\ref{HO2}). This $+1$ term plays a major role in the proxy-SU(3) approximation, as discussed in Section 7. The unitary transformation considered above is quite general, since it involves the spin and the orbital angular momentum.

\begin{turnpage}

\begingroup



\begin{table}[ht]
\caption{The same as Table \ref{N1in}, but for $\mathcal{N}=2$.}\label{N2in}

\begin{tabular}{ccccccccccccc}
\noalign{\smallskip}\hline\noalign{\smallskip}\\

$\ket{n_z,n_x,n_y,m_s}$ & $\ket{0,0,2,-{1\over 2}}$ &$\ket{0,0,2,{1\over 2}}$ & $\ket{0,1,1,-{1\over 2}}$ & $\ket{0,1,1,{1\over 2}}$ & $\ket{0,2,0,-{1\over 2}}$  & $\ket{0,2,0,{1\over 2}}$ & $\ket{1,0,1,-{1\over 2}}$ & 
$\ket{1,0,1,{1\over 2}}$ &
$\ket{1,1,0,-{1\over 2}}$ & $\ket{1,1,0,{1\over 2}}$ & $\ket{2,0,0,-{1\over 2}}$ & $\ket{2,0,0,{1\over 2}}$ \\
$\ket{n,l,j,m_j}$ &&&&&\\ 
\noalign{\smallskip}\hline\noalign{\smallskip}

$\ket{2s^{1/2}_{-1/2}}$ & $-\frac{1}{\sqrt{3}}$ & 0 & 0 & 0 & $-\frac{1}{\sqrt{3}}$ & 0 & 0 & 0 & 0 & 0 & $-\frac{1}{\sqrt{3}}$ & 0\smallskip\\
$\ket{2s^{1/2}_{1/2}}$ &0 & $-\frac{1}{\sqrt{3}}$ & 0 & 0 & 0 & $-\frac{1}{\sqrt{3}}$ & 0 & 0 & 0 & 0 & 0 & $-\frac{1}{\sqrt{3}}$ \smallskip\\
$\ket{1d^{3/2}_{-3/2}}$ & 0 & $\frac{1}{\sqrt{5}}$ & 0 & $i \sqrt{\frac{2}{5}}$ & 0 & $-\frac{1}{\sqrt{5}}$ & $-\frac{i}{\sqrt{10}}$ & 0 & $\frac{1}{\sqrt{10}}$ & 0 & 0 & 0 \smallskip\\
$\ket{1d^{3/2}_{-1/2}}$ &  $-\frac{1}{\sqrt{15}}$ & 0 & 0 & 0 & $-\frac{1}{\sqrt{15}}$ & 0 & 0 & $i \sqrt{\frac{3}{10}}$ & 0 & $-\sqrt{\frac{3}{10}}$ & $\frac{2}{\sqrt{15}}$ & 0  \smallskip\\
$\ket{1d^{3/2}_{1/2}}$ & 0 & $\frac{1}{\sqrt{15}}$ & 0 & 0 & 0 & $\frac{1}{\sqrt{15}}$ & $-i \sqrt{\frac{3}{10}}$ & 0 & $-\sqrt{\frac{3}{10}}$ & 0 & 0 & $-\frac{2}{\sqrt{15}}$  \smallskip\\
$\ket{1d^{3/2}_{3/2}}$ & $-\frac{1}{\sqrt{5}}$ & 0 & $i \sqrt{\frac{2}{5}}$ & 0 & $\frac{1}{\sqrt{5}}$ & 0 & 0 & $\frac{i}{\sqrt{10}}$ & 0 & $\frac{1}{\sqrt{10}}$ & 0 & 0 \smallskip\\
$\ket{1d^{5/2}_{-5/2}}$ &  $-\frac{1}{2}$ & 0 & $-\frac{i}{\sqrt{2}}$ & 0 & $\frac{1}{2}$ & 0 & 0 & 0 & 0 & 0 & 0 & 0 \smallskip\\
$\ket{1d^{5/2}_{-3/2}}$ &  0 & $-\frac{1}{2 \sqrt{5}}$ & 0 & $-\frac{i}{\sqrt{10}}$ & 0 & $\frac{1}{2 \sqrt{5}}$ & $-i \sqrt{\frac{2}{5}}$ & 0 & $\sqrt{\frac{2}{5}}$ & 0 & 0 & 0  \smallskip\\
$\ket{1d^{5/2}_{-1/2}}$ &$-\frac{1}{\sqrt{10}}$ & 0 & 0 & 0 & $-\frac{1}{\sqrt{10}}$ & 0 & 0 & $-\frac{i}{\sqrt{5}}$ & 0 & $\frac{1}{\sqrt{5}}$ & $\sqrt{\frac{2}{5}}$ & 0 \smallskip\\
$\ket{1d^{5/2}_{1/2}}$ & 0 & $-\frac{1}{\sqrt{10}}$ & 0 & 0 & 0 & $-\frac{1}{\sqrt{10}}$ & $-\frac{i}{\sqrt{5}}$ & 0 & $-\frac{1}{\sqrt{5}}$ & 0 & 0 & $\sqrt{\frac{2}{5}}$ \smallskip\\
$\ket{1d^{5/2}_{3/2}}$ & $-\frac{1}{2 \sqrt{5}}$ & 0 & $\frac{i}{\sqrt{10}}$ & 0 & $\frac{1}{2 \sqrt{5}}$ & 0 & 0 & $-i \sqrt{\frac{2}{5}}$ & 0 & $-\sqrt{\frac{2}{5}}$ & 0 & 0\smallskip\\
$\ket{1d^{5/2}_{5/2}}$ & 0 & $-\frac{1}{2}$ & 0 & $\frac{i}{\sqrt{2}}$ & 0 & $\frac{1}{2}$ & 0 & 0 & 0 & 0 & 0 & 0\\

\noalign{\smallskip}\hline

\end{tabular}

\end{table}

\endgroup

\end{turnpage}

\clearpage

\begin{turnpage}

\begingroup

\squeezetable 

\begin{table}
\caption{The same as Table \ref{N1}, but for $\mathcal{N}=3$, related to the harmonic oscillator shell 20-40 ($pf$ shell), or to the proxy-SU(3) shell 28-48. }\label{N3}

\begin{tabular}{ccccccccccccccccccccc}
\noalign{\smallskip}\hline\noalign{\smallskip}

$\ket{n,l,j,m_j}$ &$\ket{2p^{1/2}_{-1/2}}$ &$\ket{2p^{1/2}_{1/2}}$ &$\ket{2p^{3/2}_{-3/2}}$ &$\ket{2p^{3/2}_{-1/2}}$ &$\ket{2p^{3/2}_{1/2}}$ & $\ket{2p^{3/2}_{3/2}}$ &$\ket{1f^{5/2}_{-5/2}}$ &$\ket{1f^{5/2}_{-3/2}}$ &
$\ket{1f^{5/2}_{-1/2}}$ & $\ket{1f^{5/2}_{1/2}}$ &$\ket{1f^{5/2}_{3/2}}$ &$\ket{1f^{5/2}_{5/2}}$ &$\ket{1f^{7/2}_{-7/2}}$ &$\ket{1f^{7/2}_{-5/2}}$ &$\ket{1f^{7/2}_{-3/2}}$ &$\ket{1f^{7/2}_{-1/2}}$ &
$\ket{1f^{7/2}_{1/2}}$ &$\ket{1f^{7/2}_{3/2}}$ &$\ket{1f^{7/2}_{5/2}}$ &$\ket{1f^{7/2}_{7/2}}$ \\
$\ket{n_z,n_x,n_y,m_s}$ &&&&&\\

\noalign{\smallskip}\hline\noalign{\smallskip}

$\ket{0,0,3,-{1\over 2}}$& 0 & $-\frac{i}{\sqrt{5}}$ & $-i \sqrt{\frac{3}{10}}$ & 0 & $-\frac{i}{\sqrt{10}}$ & 0 & 0 & $-\frac{1}{2} i \sqrt{\frac{3}{35}}$ & 0 & $-i \sqrt{\frac{3}{70}}$
& 0 & $-\frac{1}{2} i \sqrt{\frac{3}{7}}$ & $-\frac{i}{2 \sqrt{2}}$ & 0 & $-\frac{1}{2} i \sqrt{\frac{3}{14}}$ & 0 & $-\frac{3 i}{2 \sqrt{70}}$ & 0 & $-\frac{i}{2 \sqrt{14}}$ & 0 \\
$\ket{0,0,3,{1\over 2}}$ & $\frac{i}{\sqrt{5}}$ & 0 & 0 & $-\frac{i}{\sqrt{10}}$ & 0 & $-i \sqrt{\frac{3}{10}}$ & $\frac{1}{2} i \sqrt{\frac{3}{7}}$ & 0 & $i \sqrt{\frac{3}{70}}$ &
0 & $\frac{1}{2} i \sqrt{\frac{3}{35}}$ & 0 & 0 & $-\frac{i}{2 \sqrt{14}}$ & 0 & $-\frac{3 i}{2 \sqrt{70}}$ & 0 & $-\frac{1}{2} i \sqrt{\frac{3}{14}}$ &
0 & $-\frac{i}{2 \sqrt{2}}$ \\
$\ket{0,1,2,-{1\over 2}}$ & 0 & $\frac{1}{\sqrt{15}}$ & $-\frac{1}{\sqrt{10}}$ & 0 & $\frac{1}{\sqrt{30}}$ & 0 & 0 & $-\frac{1}{2 \sqrt{35}}$ & 0 & $\frac{1}{\sqrt{70}}$ & 0 & $\frac{3}{2
\sqrt{7}}$ & $-\frac{\sqrt{\frac{3}{2}}}{2}$ & 0 & $-\frac{1}{2 \sqrt{14}}$ & 0 & $\frac{\sqrt{\frac{3}{70}}}{2}$ & 0 & $\frac{\sqrt{\frac{3}{14}}}{2}$ &
0 \\
$\ket{0,1,2,{1\over 2}}$ & $\frac{1}{\sqrt{15}}$ & 0 & 0 & $-\frac{1}{\sqrt{30}}$ & 0 & $\frac{1}{\sqrt{10}}$ & $\frac{3}{2 \sqrt{7}}$ & 0 & $\frac{1}{\sqrt{70}}$ & 0 & $-\frac{1}{2
\sqrt{35}}$ & 0 & 0 & $-\frac{\sqrt{\frac{3}{14}}}{2}$ & 0 & $-\frac{\sqrt{\frac{3}{70}}}{2}$ & 0 & $\frac{1}{2 \sqrt{14}}$ & 0 & $\frac{\sqrt{\frac{3}{2}}}{2}$
\\
$\ket{0,2,1,-{1\over 2}}$ & 0 & $-\frac{i}{\sqrt{15}}$ & $-\frac{i}{\sqrt{10}}$ & 0 & $-\frac{i}{\sqrt{30}}$ & 0 & 0 & $-\frac{i}{2 \sqrt{35}}$ & 0 & $-\frac{i}{\sqrt{70}}$ & 0 & $\frac{3
i}{2 \sqrt{7}}$ & $\frac{1}{2} i \sqrt{\frac{3}{2}}$ & 0 & $-\frac{i}{2 \sqrt{14}}$ & 0 & $-\frac{1}{2} i \sqrt{\frac{3}{70}}$ & 0 & $\frac{1}{2} i \sqrt{\frac{3}{14}}$
& 0 \\
$\ket{0,2,1,{1\over 2}}$ & $\frac{i}{\sqrt{15}}$ & 0 & 0 & $-\frac{i}{\sqrt{30}}$ & 0 & $-\frac{i}{\sqrt{10}}$ & $-\frac{3 i}{2 \sqrt{7}}$ & 0 & $\frac{i}{\sqrt{70}}$ & 0 & $\frac{i}{2
\sqrt{35}}$ & 0 & 0 & $\frac{1}{2} i \sqrt{\frac{3}{14}}$ & 0 & $-\frac{1}{2} i \sqrt{\frac{3}{70}}$ & 0 & $-\frac{i}{2 \sqrt{14}}$ & 0 & $\frac{1}{2} i
\sqrt{\frac{3}{2}}$ \\
$\ket{0,3,0,-{1\over 2}}$ & 0 & $\frac{1}{\sqrt{5}}$  & $-\sqrt{\frac{3}{10}}$  & 0 & $\frac{1}{\sqrt{10}}$ & 0 & 0 & $-\frac{\sqrt{\frac{3}{35}}}{2}$ & 0 & $\sqrt{\frac{3}{70}}$ & 0 &
$-\frac{\sqrt{\frac{3}{7}}}{2}$ & $\frac{1}{2 \sqrt{2}}$ & 0 & $-\frac{\sqrt{\frac{3}{14}}}{2}$ & 0 & $\frac{3}{2 \sqrt{70}}$ & 0 & $-\frac{1}{2 \sqrt{14}}$
& 0 \\
$\ket{0,3,0,{1\over 2}}$& $\frac{1}{\sqrt{5}}$ & 0 & 0 & $-\frac{1}{\sqrt{10}}$ & 0 & $\sqrt{\frac{3}{10}}$ & $-\frac{\sqrt{\frac{3}{7}}}{2}$ & 0 & $\sqrt{\frac{3}{70}}$ & 0 & $-\frac{\sqrt{\frac{3}{35}}}{2}$
& 0 & 0 & $\frac{1}{2 \sqrt{14}}$ & 0 & $-\frac{3}{2 \sqrt{70}}$ & 0 & $\frac{\sqrt{\frac{3}{14}}}{2}$ & 0 & $-\frac{1}{2 \sqrt{2}}$ \\
$\ket{1,0,2,-{1\over 2}}$& $-\frac{1}{\sqrt{15}}$ & 0 & 0 & $-\sqrt{\frac{2}{15}}$ & 0 & 0 & $-\frac{1}{2 \sqrt{7}}$ & 0 & $-\frac{3}{\sqrt{70}}$ & 0 & $-\frac{\sqrt{\frac{5}{7}}}{2}$
& 0 & 0 & $-\sqrt{\frac{3}{14}}$ & 0 & $-\sqrt{\frac{6}{35}}$ & 0 & $-\frac{1}{\sqrt{14}}$ & 0 & 0 \\
$\ket{1,0,2,{1\over 2}}$ & 0 & $\frac{1}{\sqrt{15}}$ & 0 & 0 & $-\sqrt{\frac{2}{15}}$ & 0 & 0 & $\frac{\sqrt{\frac{5}{7}}}{2}$ & 0 & $\frac{3}{\sqrt{70}}$ & 0 & $\frac{1}{2 \sqrt{7}}$
& 0 & 0 & $-\frac{1}{\sqrt{14}}$ & 0 & $-\sqrt{\frac{6}{35}}$ & 0 & $-\sqrt{\frac{3}{14}}$ & 0 \\
$\ket{1,1,1,-{1\over 2}}$& 0 & 0 & 0 & 0 & 0 & 0 & $\frac{i}{\sqrt{14}}$ & 0 & 0 & 0 & $-i \sqrt{\frac{5}{14}}$ & 0 & 0 & $i \sqrt{\frac{3}{7}}$ & 0 & 0 & 0 & $-\frac{i}{\sqrt{7}}$
& 0 & 0 \\
$\ket{1,1,1,{1\over 2}}$ & 0 & 0 & 0 & 0 & 0 & 0 & 0 & $-i \sqrt{\frac{5}{14}}$ & 0 & 0 & 0 & $\frac{i}{\sqrt{14}}$ & 0 & 0 & $\frac{i}{\sqrt{7}}$ & 0 & 0 & 0 & $-i \sqrt{\frac{3}{7}}$
& 0 \\
$\ket{1,2,0,-{1\over 2}}$ & $-\frac{1}{\sqrt{15}}$ & 0 & 0 & $-\sqrt{\frac{2}{15}}$ & 0 & 0 & $\frac{1}{2 \sqrt{7}}$ & 0 & $-\frac{3}{\sqrt{70}}$ & 0 & $\frac{\sqrt{\frac{5}{7}}}{2}$
& 0 & 0 & $\sqrt{\frac{3}{14}}$ & 0 & $-\sqrt{\frac{6}{35}}$ & 0 & $\frac{1}{\sqrt{14}}$ & 0 & 0 \\
$\ket{1,2,0,{1\over 2}}$& 0 & $\frac{1}{\sqrt{15}}$ & 0 & 0 & $-\sqrt{\frac{2}{15}}$ & 0 & 0 & $-\frac{\sqrt{\frac{5}{7}}}{2}$ & 0 & $\frac{3}{\sqrt{70}}$ & 0 & $-\frac{1}{2 \sqrt{7}}$
& 0 & 0 & $\frac{1}{\sqrt{14}}$ & 0 & $-\sqrt{\frac{6}{35}}$ & 0 & $\sqrt{\frac{3}{14}}$ & 0 \\
$\ket{2,0,1,-{1\over 2}}$ & 0 & $-\frac{i}{\sqrt{15}}$ & $-\frac{i}{\sqrt{10}}$ & 0 & $-\frac{i}{\sqrt{30}}$ & 0 & 0 & $\frac{2 i}{\sqrt{35}}$ & 0 & $2 i \sqrt{\frac{2}{35}}$ & 0 & 0
& 0 & 0 & $i \sqrt{\frac{2}{7}}$ & 0 & $i \sqrt{\frac{6}{35}}$ & 0 & 0 & 0 \\
$\ket{2,0,1,{1\over 2}}$ & $\frac{i}{\sqrt{15}}$ & 0 & 0 & $-\frac{i}{\sqrt{30}}$ & 0 & $-\frac{i}{\sqrt{10}}$ & 0 & 0 & $-2 i \sqrt{\frac{2}{35}}$ & 0 & $-\frac{2 i}{\sqrt{35}}$ &
0 & 0 & 0 & 0 & $i \sqrt{\frac{6}{35}}$ & 0 & $i \sqrt{\frac{2}{7}}$ & 0 & 0 \\
$\ket{2,1,0,-{1\over 2}}$ & 0 & $\frac{1}{\sqrt{15}}$ & $-\frac{1}{\sqrt{10}}$ & 0 & $\frac{1}{\sqrt{30}}$ & 0 & 0 & $\frac{2}{\sqrt{35}}$ & 0 & $-2 \sqrt{\frac{2}{35}}$ & 0 & 0 & 0
& 0 & $\sqrt{\frac{2}{7}}$ & 0 & $-\sqrt{\frac{6}{35}}$ & 0 & 0 & 0 \\
$\ket{2,1,0,{1\over 2}}$ & $\frac{1}{\sqrt{15}}$ & 0 & 0 & $-\frac{1}{\sqrt{30}}$ & 0 & $\frac{1}{\sqrt{10}}$ & 0 & 0 & $-2 \sqrt{\frac{2}{35}}$ & 0 & $\frac{2}{\sqrt{35}}$ & 0 & 0
& 0 & 0 & $\sqrt{\frac{6}{35}}$ & 0 &$ -\sqrt{\frac{2}{7}}$ & 0 & 0 \\
$\ket{3,0,0,-{1\over 2}}$ & $-\frac{1}{\sqrt{5}}$ & 0 & 0 & $ -\sqrt{\frac{2}{5}}$ & 0 & 0 & 0 & 0 & $\sqrt{\frac{6}{35}}$ & 0 & 0 & 0 & 0 & 0 & 0 & $2 \sqrt{\frac{2}{35}}$ & 0 & 0
& 0 & 0 \\
$\ket{3,0,0,{1\over 2}}$ & 0 & $\frac{1}{\sqrt{5}}$ & 0 & 0 & $-\sqrt{\frac{2}{5}}$ & 0 & 0 & 0 & 0 & $-\sqrt{\frac{6}{35}}$ & 0 & 0 & 0 & 0 & 0 & 0 & $2 \sqrt{\frac{2}{35}}$ & 0
& 0 & 0 \\

\noalign{\smallskip}\hline
\end{tabular}

\end{table}

\endgroup

\end{turnpage}

\clearpage 

\begin{turnpage}

\begingroup

\squeezetable
\begin{table}
\caption{The same as Table \ref{N1in}, but for $\mathcal{N}=3$.}\label{N3in}

\begin{tabular}{cccccccccccccccccccccc}
\noalign{\smallskip}\hline\noalign{\smallskip}

$\ket{n_z,n_x,n_y,m_s}$ & $\ket{0,0,3,-{1\over 2}}$ & $\ket{0,0,3,{1\over 2}}$ & $\ket{0,1,2,-{1\over 2}}$ & $\ket{0,1,2,{1\over 2}}$ &$\ket{0,2,1,-{1\over 2}}$ &$\ket{0,2,1,{1\over 2}}$ &$\ket{0,3,0,-{1\over 2}}$ &
$\ket{0,3,0,{1\over 2}}$ & 
$\ket{1,0,2,-{1\over 2}}$ & $\ket{1,0,2,{1\over 2}}$ & $\ket{1,1,1,-{1\over 2}}$ &$\ket{1,1,1,{1\over 2}}$ &$\ket{1,2,0,-{1\over 2}}$ &$\ket{1,2,0,{1\over 2}}$ & $\ket{2,0,1,-{1\over 2}}$&$\ket{2,0,1,{1\over 2}}$ & 
$\ket{2,1,0,-{1\over 2}}$& 
$\ket{2,1,0,{1\over 2}}$ & $\ket{3,0,0,-{1\over 2}}$ & $\ket{3,0,0,{1\over 2}}$ \\
$\ket{n,l,j,m_j}$ \\

\noalign{\smallskip}\hline\noalign{\smallskip}

$\ket{2p^{1/2}_{-1/2}}$ & 0 & $ -\frac{i}{\sqrt{5}}$ & 0 & $\frac{1}{\sqrt{15}}$ & 0 & $-\frac{i}{\sqrt{15}}$ & 0 & $\frac{1}{\sqrt{5}}$ & $-\frac{1}{\sqrt{15}}$ & 0 & 0 & 0 & $-\frac{1}{\sqrt{15}}$
& 0 & 0 & $-\frac{i}{\sqrt{15}}$ & 0 & $\frac{1}{\sqrt{15}}$ & $-\frac{1}{\sqrt{5}}$ & 0 \\
$\ket{2p^{1/2}_{1/2}}$ & $\frac{i}{\sqrt{5}}$ & 0 & $\frac{1}{\sqrt{15}}$ & 0 & $\frac{i}{\sqrt{15}}$ & 0 & $\frac{1}{\sqrt{5}}$ & 0 & 0 & $\frac{1}{\sqrt{15}}$ & 0 & 0 & 0 & $\frac{1}{\sqrt{15}}$
& $\frac{i}{\sqrt{15}}$ & 0 & $\frac{1}{\sqrt{15}}$ & 0 & 0 & $\frac{1}{\sqrt{5}}$ \\
$\ket{2p^{3/2}_{-3/2}}$ & $i \sqrt{\frac{3}{10}}$ & 0 & $-\frac{1}{\sqrt{10}}$ & 0 & $\frac{i}{\sqrt{10}}$ & 0 & $-\sqrt{\frac{3}{10}}$ & 0 & 0 & 0 & 0 & 0 & 0 & 0 & $\frac{i}{\sqrt{10}}$
& 0 & $-\frac{1}{\sqrt{10}}$ & 0 & 0 & 0 \\
$\ket{2p^{3/2}_{-1/2}}$ & 0 & $\frac{i}{\sqrt{10}}$ & 0 & $-\frac{1}{\sqrt{30}}$ & 0 & $\frac{i}{\sqrt{30}}$ & 0 & $-\frac{1}{\sqrt{10}}$ & $-\sqrt{\frac{2}{15}}$ & 0 & 0 & 0 & $-\sqrt{\frac{2}{15}}$
& 0 & 0 & $\frac{i}{\sqrt{30}}$ & 0 & $-\frac{1}{\sqrt{30}}$ & $-\sqrt{\frac{2}{5}}$ & 0 \\
$\ket{2p^{3/2}_{1/2}}$ & $\frac{i}{\sqrt{10}}$ & 0 & $\frac{1}{\sqrt{30}}$ & 0 & $\frac{i}{\sqrt{30}}$ & 0 & $\frac{1}{\sqrt{10}}$ & 0 & 0 & $-\sqrt{\frac{2}{15}}$ & 0 & 0 & 0 & $-\sqrt{\frac{2}{15}}$
& $\frac{i}{\sqrt{30}}$ & 0 & $\frac{1}{\sqrt{30}}$ & 0 & 0 & $-\sqrt{\frac{2}{5}}$ \\
$\ket{2p^{3/2}_{3/2}}$& 0 & $i \sqrt{\frac{3}{10}}$ & 0 & $\frac{1}{\sqrt{10}}$ & 0 & $\frac{i}{\sqrt{10}}$ & 0 & $\sqrt{\frac{3}{10}}$ & 0 & 0 & 0 & 0 & 0 & 0 & 0 & $\frac{i}{\sqrt{10}}$
& 0 & $\frac{1}{\sqrt{10}}$ & 0 & 0 \\
$\ket{1f^{5/2}_{-5/2}}$& 0 & $-\frac{1}{2} i \sqrt{\frac{3}{7}}$ & 0 & $\frac{3}{2 \sqrt{7}}$ & 0 & $\frac{3 i}{2 \sqrt{7}}$ & 0 & $-\frac{\sqrt{\frac{3}{7}}}{2}$ & $-\frac{1}{2
\sqrt{7}}$ & 0 & $-\frac{i}{\sqrt{14}}$ & 0 & $\frac{1}{2 \sqrt{7}}$ & 0 & 0 & 0 & 0 & 0 & 0 & 0 \\
$\ket{1f^{5/2}_{-3/2}}$ & $\frac{1}{2} i \sqrt{\frac{3}{35}}$ & 0 & $-\frac{1}{2 \sqrt{35}}$ & 0 & $\frac{i}{2 \sqrt{35}}$ & 0 & $-\frac{\sqrt{\frac{3}{35}}}{2}$ & 0 & 0 & $\frac{\sqrt{\frac{5}{7}}}{2}$
& 0 & $i \sqrt{\frac{5}{14}}$ & 0 & $-\frac{\sqrt{\frac{5}{7}}}{2}$ & $-\frac{2 i}{\sqrt{35}}$ & 0 & $\frac{2}{\sqrt{35}}$ & 0 & 0 & 0 \\
$\ket{1f^{5/2}_{-1/2}}$ & 0 & $-i \sqrt{\frac{3}{70}}$ & 0 & $\frac{1}{\sqrt{70}}$ & 0 & $-\frac{i}{\sqrt{70}}$ & 0 & $\sqrt{\frac{3}{70}}$ & $-\frac{3}{\sqrt{70}}$ & 0 & 0 & 0 & $-\frac{3}{\sqrt{70}}$
& 0 & 0 & $2 i \sqrt{\frac{2}{35}}$ & 0 & $-2 \sqrt{\frac{2}{35}}$ & $\sqrt{\frac{6}{35}}$ & 0 \\
$\ket{1f^{5/2}_{1/2}}$ & $i \sqrt{\frac{3}{70}}$ & 0 & $\frac{1}{\sqrt{70}}$ & 0 & $\frac{i}{\sqrt{70}}$ & 0 & $\sqrt{\frac{3}{70}}$ & 0 & 0 & $\frac{3}{\sqrt{70}}$ & 0 & 0 & 0 &
$\frac{3}{\sqrt{70}}$ & $-2 i \sqrt{\frac{2}{35}}$ & 0 & $-2 \sqrt{\frac{2}{35}}$ & 0 & 0 & $-\sqrt{\frac{6}{35}}$ \\
$\ket{1f^{5/2}_{3/2}}$& 0 & $-\frac{1}{2} i \sqrt{\frac{3}{35}}$ & 0 & $-\frac{1}{2 \sqrt{35}}$ & 0 & $-\frac{i}{2 \sqrt{35}}$ & 0 & $-\frac{\sqrt{\frac{3}{35}}}{2}$ & $-\frac{\sqrt{\frac{5}{7}}}{2}$
& 0 & $i \sqrt{\frac{5}{14}}$ & 0 & $\frac{\sqrt{\frac{5}{7}}}{2}$ & 0 & 0 & $\frac{2 i}{\sqrt{35}}$ & 0 & $\frac{2}{\sqrt{35}}$ & 0 & 0 \\
$\ket{1f^{5/2}_{5/2}}$ & $\frac{1}{2} i \sqrt{\frac{3}{7}}$ & 0 & $\frac{3}{2 \sqrt{7}}$ & 0 & $-\frac{3 i}{2 \sqrt{7}}$ & 0 & $-\frac{\sqrt{\frac{3}{7}}}{2}$ & 0 & 0 & $\frac{1}{2\sqrt{7}}$
 & 0 & $-\frac{i}{\sqrt{14}}$ & 0 & $-\frac{1}{2 \sqrt{7}}$ & 0 & 0 & 0 & 0 & 0 & 0 \\
$\ket{1f^{7/2}_{-7/2}}$ & $\frac{i}{2 \sqrt{2}}$ & 0 & $-\frac{\sqrt{\frac{3}{2}}}{2}$ & 0 & $-\frac{1}{2} i \sqrt{\frac{3}{2}}$ & 0 & $\frac{1}{2 \sqrt{2}}$ & 0 & 0 & 0 & 0 & 0
& 0 & 0 & 0 & 0 & 0 & 0 & 0 & 0 \\
$\ket{1f^{7/2}_{-5/2}}$ & 0 & $\frac{i}{2 \sqrt{14}}$ & 0 & $-\frac{\sqrt{\frac{3}{14}}}{2}$ & 0 & $-\frac{1}{2} i \sqrt{\frac{3}{14}}$ & 0 & $\frac{1}{2 \sqrt{14}}$ & $-\sqrt{\frac{3}{14}}$
& 0 & $-i \sqrt{\frac{3}{7}}$ & 0 & $\sqrt{\frac{3}{14}}$ & 0 & 0 & 0 & 0 & 0 & 0 & 0 \\
$\ket{1f^{7/2}_{-3/2}}$& $\frac{1}{2} i \sqrt{\frac{3}{14}}$ & 0 & $-\frac{1}{2 \sqrt{14}}$ & 0 & $\frac{i}{2 \sqrt{14}}$ & 0 & $-\frac{\sqrt{\frac{3}{14}}}{2}$ & 0 & 0 & $-\frac{1}{\sqrt{14}}$
& 0 & $-\frac{i}{\sqrt{7}}$ & 0 & $\frac{1}{\sqrt{14}}$ & $-i \sqrt{\frac{2}{7}}$ & 0 & $\sqrt{\frac{2}{7}}$ & 0 & 0 & 0 \\
$\ket{1f^{7/2}_{-1/2}}$ &0 & $\frac{3 i}{2 \sqrt{70}}$ & 0 & $-\frac{\sqrt{\frac{3}{70}}}{2}$ & 0 & $\frac{1}{2} i \sqrt{\frac{3}{70}}$ & 0 & $-\frac{3}{2 \sqrt{70}}$ & $-\sqrt{\frac{6}{35}}$
& 0 & 0 & 0 & $-\sqrt{\frac{6}{35}}$ & 0 & 0 & $-i \sqrt{\frac{6}{35}}$ & 0 & $\sqrt{\frac{6}{35}}$ & $2 \sqrt{\frac{2}{35}}$ & 0 \\
$\ket{1f^{7/2}_{1/2}}$ & $\frac{3 i}{2 \sqrt{70}}$ & 0 & $\frac{\sqrt{\frac{3}{70}}}{2}$ & 0 & $\frac{1}{2} i \sqrt{\frac{3}{70}}$ & 0 & $\frac{3}{2 \sqrt{70}}$ & 0 & 0 & $-\sqrt{\frac{6}{35}}$
& 0 & 0 & 0 & $-\sqrt{\frac{6}{35}}$ & $-i \sqrt{\frac{6}{35}}$ & 0 & $-\sqrt{\frac{6}{35}}$ & 0 & 0 & $2 \sqrt{\frac{2}{35}}$ \\
$\ket{1f^{7/2}_{3/2}}$ & 0 & $\frac{1}{2} i \sqrt{\frac{3}{14}}$ & 0 & $\frac{1}{2 \sqrt{14}}$ & 0 & $\frac{i}{2 \sqrt{14}}$ & 0 & $\frac{\sqrt{\frac{3}{14}}}{2}$ & $-\frac{1}{\sqrt{14}}$
& 0 & $\frac{i}{\sqrt{7}}$ & 0 & $\frac{1}{\sqrt{14}}$ & 0 & 0 & $-i \sqrt{\frac{2}{7}}$ & 0 & $-\sqrt{\frac{2}{7}}$ & 0 & 0 \\
$\ket{1f^{7/2}_{5/2}}$ & $\frac{i}{2 \sqrt{14}}$ & 0 & $\frac{\sqrt{\frac{3}{14}}}{2}$ & 0 & $-\frac{1}{2} i \sqrt{\frac{3}{14}}$ & 0 & $-\frac{1}{2 \sqrt{14}}$ & 0 & 0 & $-\sqrt{\frac{3}{14}}$
& 0 & $i \sqrt{\frac{3}{7}}$ & 0 & $\sqrt{\frac{3}{14}}$ & 0 & 0 & 0 & 0 & 0 & 0 \\
$\ket{1f^{7/2}_{7/2}}$ & 0 & $\frac{i}{2 \sqrt{2}}$ & 0 & $\frac{\sqrt{\frac{3}{2}}}{2}$ & 0 & $-\frac{1}{2} i \sqrt{\frac{3}{2}}$ & 0 & $-\frac{1}{2 \sqrt{2}}$ & 0 & 0 & 0 & 0
& 0 & 0 & 0 & 0 & 0 & 0 & 0 & 0 \\

\noalign{\smallskip}\hline
\end{tabular}

\end{table}
\endgroup

\end{turnpage}

\clearpage


\begin{table}[htb]
\caption{The shell model orbitals of the original spin-orbit like shells and of the proxy-SU(3) shells. The magic number $14$ is proposed as a sub-shell closure in Ref. \cite{Sorlin}. The symmetry of each proxy-SU(3) shell is U($\Omega$) with $\Omega={(\mathcal{N}+1)(\mathcal{N}+2)\over 2}$. The orbitals, that are being replaced, are denoted with bold letters. See Section 5 for further discussion.}\label{shells}
\begin{tabular}{cccccc}
\noalign{\smallskip}\hline\noalign{\smallskip}
spin-orbit    &                    &                &                            & proxy-SU(3)    &  3D-HO        \\          
magic numbers & original orbitals & proxy orbitals &proxy  $U(\Omega)$ symmetry &  magic numbers & magic numbers \\
 \noalign{\smallskip}\hline\noalign{\smallskip}
 
6-14 & $1p^{1/2}_{\pm 1/2}$ & $1p^{1/2}_{\pm 1/2}$ & $U(3)$ & 6-12 & 2-8 \smallskip\\
&  $\bf 1d^{5/2}_{\pm 1/2,\pm 3/2}$&  $\bf 1p^{3/2}_{\pm 1/2,\pm 3/2}$ & & &  \smallskip\\
 & $\bf 1d^{5/2}_{\pm 5/2}$ & - & & \bigskip\\
 
 14-28 & $2s^{1/2}_{\pm 1/2}$ & $2s^{1/2}_{\pm 1/2}$ &  $U(6)$ & 14-26 & 8-20 \smallskip\\
 & $1d^{3/2}_{\pm 1/2,\pm 3/2}$ & $1d^{3/2}_{\pm 1/2,\pm 3/2}$ & & & \smallskip\\
 & $\bf 1f^{7/2}_{\pm 1/2,\pm 3/2,\pm 5/2}$ & $\bf 1d^{5/2}_{\pm 1/2,\pm 3/2,\pm 5/2}$ & & & \smallskip\\
& $\bf 1f^{7/2}_{\pm 7/2}$ & - & & & \bigskip\\

28-50 & $2p^{1/2}_{\pm 1/2}$ & $2p^{1/2}_{\pm 1/2}$ &  $U(10)$ & 28-48 & 20-40 \smallskip\\
& $2p^{3/2}_{\pm 1/2,\pm 3/2}$ &  $2p^{3/2}_{\pm 1/2,\pm 3/2}$ & & & \smallskip\\
& $1f^{5/2}_{\pm 5/2, \pm3/2,\pm 1/2}$ & $1f^{5/2}_{\pm 5/2, \pm3/2,\pm 1/2}$ & & & \smallskip\\
& $\bf 1g^{9/2}_{\pm 1/2,..., \pm 7/2}$ & $\bf 1f^{7/2}_{\pm 1/2,..., \pm 7/2}$ & & & \smallskip\\
& $\bf 1g^{9/2}_{\pm 9/2}$ & - & & & \bigskip\\

50-82  & $3s^{1/2}_{\pm 1/2}$ &  $3s^{1/2}_{\pm 1/2}$  & $U(15)$ & 50-80 & 40-70 \smallskip\\
 & $2d^{3/2}_{\pm 1/2,\pm 3/2}$ & $2d^{3/2}_{\pm 1/2,\pm 3/2}$ & & & \smallskip\\
& $2d^{5/2}_{\pm 1/2,...,\pm 5/2}$ &  $2d^{5/2}_{\pm 1/2,...,\pm 5/2}$ & & & \smallskip\\
 & $1g^{7/2}_{\pm 1/2,...,\pm 7/2}$ & $1g^{7/2}_{\pm 1/2,...,\pm 7/2}$ & & & \smallskip\\
 & $\bf 1h^{11/2}_{\pm 1/2,...,\pm 9/2}$ & $\bf 1g^{9/2}_{\pm 1/2,...,\pm 9/2}$ & & & \smallskip\\
 & $\bf 1h^{11/2}_{\pm 11/2}$ & - & & & \bigskip\\
 
82-126 & $3p^{1/2}_{\pm 1/2}$ & $3p^{1/2}_{\pm 1/2}$ & $U(21)$ & 82-124 & 70-112 \smallskip\\
 & $3p^{3/2}_{\pm 1/2,\pm 3/2}$ & $3p^{3/2}_{\pm 1/2,\pm 3/2}$ & & & \smallskip\\
& $2f^{5/2}_{\pm 1/2,...,\pm 5/2}$ & $2f^{5/2}_{\pm 1/2,...,\pm 5/2}$ & & & \smallskip\\
& $2f^{7/2}_{\pm 1/2,...,\pm 7/2}$ & $2f^{7/2}_{\pm 1/2,...,\pm 7/2}$ & & & \smallskip\\
 & $1h^{9/2}_{\pm 1/2,...,\pm 9/2}$ & $1h^{9/2}_{\pm 1/2,...,\pm 9/2}$ & & & \smallskip\\
 & $\bf 1i^{13/2}_{\pm 1/2,...,\pm 11/2}$ & $\bf 1h^{11/2}_{\pm 1/2,...,\pm 11/2}$ & & & \smallskip\\
 & $\bf 1i^{13/2}_{\pm 13/2}$ & - & & & \bigskip \\
 
126-184 & $4s^{1/2}_{\pm 1/2}$ & $4s^{1/2}_{\pm 1/2}$  & $U(28)$ & 126-182 & 112-168 \smallskip\\
& $3d^{3/2}_{\pm 1/2,\pm 3/2}$ &  $3d^{3/2}_{\pm 1/2,\pm 3/2}$ & & & \smallskip\\
& $3d^{5/2}_{\pm 1/2,...,\pm 5/2}$ & $3d^{5/2}_{\pm 1/2,...,\pm 5/2}$ & & \smallskip\\
& $2g^{7/2}_{\pm 1/2,...,\pm 7/2}$ &  $2g^{7/2}_{\pm 1/2,...,\pm 7/2}$& & & \smallskip\\
& $2g^{9/2}_{\pm 1/2,...,\pm 9/2}$ &  $2g^{9/2}_{\pm 1/2,...,\pm 9/2}$ & & & \smallskip\\
& $1i^{11/2}_{\pm 1/2,...,\pm 11/2}$ & $1i^{11/2}_{\pm 1/2,...,\pm 11/2}$ & & & \smallskip\\
& $\bf 1j^{15/2}_{\pm 1/2,...,\pm 13/2}$ & $\bf 1i^{13/2}_{\pm 1/2,...,\pm 13/2}$ & & & \smallskip\\
& $\bf 1j^{15/2}_{\pm 15/2}$ & - & & & \bigskip\\

\noalign{\smallskip}\hline\noalign{\smallskip}
\end{tabular}
\end{table}

\end{document}